\documentclass[12pt]{emulateapj}
\usepackage{epsfig}

\def\spose#1{\hbox to 0pt{#1\hss}}

\let\approxlt=\lesssim

\def\multleft#1{\hbox to size{\vbox {\halign {\lft{##}\cr #1}}\hfill}\par}
\def\multright#1{\hbox to size{\vbox {\halign {\rt{##}\cr #1}}\hfill}\par}

\def\boxit#1{\vbox{\hrule\hbox{\vrule\kern3pt\vbox{\kern3pt
          #1 \kern3pt}\kern3pt\vrule}\hrule}}

\def\Hz{{\rm\thinspace Hz}}



\begin{document}

\title{On the time variability of geometrically-thin black hole
accretion disks I : the search for modes in simulated disks.}

\author{Christopher~S.~Reynolds\altaffilmark{1} and 
M.~Coleman~Miller\altaffilmark{1}}

\altaffiltext{1}{Department of Astronomy and the Maryland Astronomy
Center for Theory and Computation, University of Maryland, College
Park, MD 20742-2421}

\begin{abstract}
We present a detailed temporal analysis of a set of hydrodynamic and
magnetohydrodynamic (MHD) simulations of geometrically-thin ($h/r\sim
0.05$) black hole accretion disks.  The black hole potential is
approximated by the Paczynski-Wiita pseudo-Newtonian potential.  In
particular, we use our simulations to critically assess two widely
discussed models for high-frequency quasi-periodic oscillations,
global oscillation modes (diskoseismology) and parametric resonance
instabilities.  We find that initially disturbed hydrodynamic disks
clearly display the trapped global g-mode oscillation predicted by
linear perturbation theory.  In contrast, the sustained turbulence
produced in the simulated MHD disks by the magneto-rotational
instability does not excite these trapped g-modes.  We cannot say at
present whether the MHD turbulence actively damps the hydrodynamic
g-mode.  Our simulated MHD disks also fail to display any indications
of a parametric resonance instability between the vertical and radial
epicyclic frequencies.  On the other hand, we do see characteristic
frequencies at any given radius in the disk corresponding to local
acoustic waves.  We also conduct a blind search for {\it any}
quasi-periodic oscillation in a proxy lightcurve based on the
instantaneous mass accretion rate of the black hole, and place an
upper limit of 2\% on the total power in any such feature. We
highlight the importance of correcting for secular changes in the
simulated accretion disk when performing temporal analyses.
\end{abstract}


\section{Introduction}

Rapid X-ray variability is a ubiquitous characteristic of accretion
onto black holes.  Aperiodic X-ray fluctuations are seen from both
Galactic Black Hole Binaries (GBHBs) and Active Galactic Nuclei (AGN)
and, accounting for the inverse scaling of all relevant frequencies
with black hole mass, appear to have similar characteristics
(Uttley, McHardy \& Vaughan 2005; McHardy et al. 2006).  While it is
highly tempting to relate this variability to the magnetohydrodynamic
(MHD) turbulence that is believed to drive the accretion process
(Balbus \& Hawley 1991, 1998), the exact physical
processes underlying the observed fluctuations remain mysterious.

GBHBs also display quasi-periodic oscillations (QPOs) in their X-ray
lightcurves (see review by McClintock \& Remillard 2003)\footnote{Very
recently, the first convincing case of a QPO in an AGN was reported by
Gierlinski et al. (2008).}.  Of particular interest are the
high-frequency quasi-periodic oscillations (HFQPOs) that are seen in
the very-high (or steep power law; McClintock \& Remillard 2003) state
of GBHBs.  The HFQPOs have quality factors of few-to-$10$, centroid
frequencies of order $100\Hz$ and appear to be imprinted on the hard
X-ray tail of the spectrum rather than the thermal disk emission.  The
fact that their frequencies are stable and at least loosely comparable
to the orbital frequency at the innermost stable circular orbit (ISCO)
around the black hole suggests that their properties are set by the
relativistic portions of the gravitational potential.  This gives them
enormous promise as a diagnostic of black hole mass and spin.

However, the utility of HFQPOs to relativistic astrophysics is
severely limited by the lack of a compelling theoretical framework in
which to interpret measurements of the frequencies, quality factors
and root-mean-square (rms) powers.  There exist well defined geodesic
frequencies (i.e., the orbital, radial epicyclic and vertical
epicyclic frequencies) at any given radius in the accretion disk.
However, these frequencies (as well as all non-trivial linear
combinations) change with radius, and it is not clear why the
frequencies of any one particular radius would be preferentially
displayed in the overall power-spectrum.  

HFQPOs are commonly found in pairs with an approximate 3:2 frequency
ratio, and this has been used to suggest that a particular radius is
picked out due to a resonance.  In the parametric resonance model
(Abramowicz \& Klu\'zniak 2001, 2003; Abramowicz et al.  2002, 2003),
there is a resonance in the disk at the radius where the radial
epicyclic frequency and vertical epicyclic frequency are in small
integer ratios.  As discussed below, the strongest resonance occurs
when these frequencies are in 3:2 ratio, at least in the simplest
manifestation of this model.  Other resonance models have been
examined by Rezzolla et al. (2003a,b) and Kato (2004a,b,c).

Another interesting possibility is that the HFQPOs are global
oscillation modes of the accretion disk (i.e., ``diskoseismic''
modes).  Global modes have been examined analytically (using linear
theory) on a hydrodynamic background in spacetimes that are
pseudo-Newtonian (Okazaki, Kato, \& Fukue 1987; Nowak \& Wagoner 1991,
1992, 1993; Markovi\'c \& Lamb 1998), Schwarzschild (Kato \& Fukue
1980), and Kerr (Kato 1990, 1991, 1993; Kato \& Honma 1991; Perez et
al. 1997; Silbergleit et al. 2001; Wagoner et al. 2001;
Ortega-Rodriguez et al. 2001).  Three classes of mode are recovered
corresponding to pressure modes (p-modes), inertial modes
(conventionally referred to as g-modes even though the restoring force
results from rotation or inertia, depending on your frame of reference;
J.Pringle priv. communication) and warping/corrugation modes
(c-modes).  For plausible masses and spins, the fundamental ($m=0$)
g-mode was quickly identified as a good candidate for the first HFQPO
discovered, the 67\,Hz oscillation found in the system GRS~1915$+$105
(Nowak et al. 1997).  Current diskoseismology theory does not provide
a natural explanation of HFQPO pairs with small-integer ratios;
however, all present analyses are conducted using linear theory
whereas these HFQPOs pairs would likely arise from mode-coupling that
would only be revealed by a non-linear analysis.

Clearly, many open questions concerning the physics of X-ray
variability remain, including the correct interpretation of HFQPOs.
The current dominant paradigm for understanding black hole accretion
is that the magnetorotational instability (MRI; Balbus \& Hawley 1991)
drives powerful MHD turbulence, and correlated Maxwell stresses within
this turbulence mediate the outward transport of angular momentum
that allows accretion to proceed.  However, the connection between the
MHD turbulence paradigm and models for the aperiodic and
quasi-periodic variability remains highly uncertain.  For example, can
the MHD turbulence naturally produce the rms-flux relation noted in
most black hole X-ray lightcurves (Uttley \& McHardy 2001) and/or the
log-normal flux distribution found in Cygnus X-1 (Uttley, McHardy \&
Vaughan 2005)?  Are diskoseismic modes excited by turbulent
fluctuation (Nowak \& Wagoner 1993), or does the turbulence act to
damp such modes (Arras, Blaes \& Turner 2006)?  Do the Maxwell
stresses couple radial and vertical motions in such a way as to excite
parametric resonance instabilities of the type identified by
Abramowicz \& Klu\'zniak (2001) or any other resonant phenomena?  Does
the fact that the HFQPOs are imprinted on the high-energy tail provide
a fundamental clue to their origin, or is it a generic consequence of
any oscillating thermal accretion disk surrounded by a Comptonizing
corona (Lehr, Wagoner \& Wilms 2000)?

In this paper, we use a set of global hydrodynamic and MHD simulations
of geometrically-thin accretion disks in a pseudo-Newtonian potential
to begin an exploration of these issues.  Our canonical MHD simulation
represents a thinner disk, and is run for more orbits, than any
previously published well-resolved 3-d MHD disk simulation.  This
allows us to conduct a more extensive study of the temporal
variability of such disks than has previously been attempted.  In
\S~2, we give a brief review of the theory of both local and global
hydrodynamic modes of black hole accretion disks, as well as the
parametric resonance instability model for HFQPOs.  \S~3 presents our
study of ideal (inviscid) hydrodynamic disks, both with imposed
axisymmetry and in full 3-dimensions.  We find prominent trapped
g-modes in the axisymmetric simulations which remain (albeit with
diminished amplitude) in the full 3-d case.  We then study the MHD
case in \S~4, where we find that the turbulence excites neither the
diskoseismic modes nor the parametric resonances discussed above.
Instead, we find that the turbulence excites local hydrodynamic waves
of the type elucidated by Lubow \& Pringle (1993).  We discuss our
results, including a comparison to previous work, in \S~5 and conclude
in \S~6.

\section{Theoretical Preliminaries}

Here we provide a brief review of some previously established
theoretical results that are pertinent to this paper.

\subsection{Local oscillations and waves in accretion disks}
\label{sec:local_osc}

There is a very extensive literature on oscillations and waves in
accretion disks.  Here we focus on just those aspects of the field
that turn out to be relevant for the interpretation of our simulation
which have been elucidated most clearly by Lubow \& Pringle (1993).

Lubow \& Pringle (1993; hereafter LP93) studied three dimensional wave
propagation in accretion disks ignoring self-gravity.  In the case
where one ignores vertical motions, they show that radial waves obey
the well-known dispersion relation
\begin{equation}
\omega^2=\kappa^2+c_s^2k^2,
\label{eq:disk_dispersion}
\end{equation}
where $c_s$ is the sound speed (assumed, in this case, to be purely a
function of $r$) and $\kappa$ is the radial epicyclic frequency given
by $\kappa^2=4\Omega^2+r\,\partial \Omega^2/\partial r$ (also see
Binney \& Tremaine 1987).  Here, $\Omega(r)$ is the angular frequency
of the background Keplerian flow.  LP93 proceed to study the
propagation of axisymmetric waves in the case where the atmosphere has
a locally isothermal vertical structure.  They find two types of
waves.  There are low-frequency gravity waves for which
\begin{equation}
0<\omega<\Omega.
\end{equation}
There are also high-frequency acoustic waves which have
\begin{equation}
\omega^2>(n\gamma+1)\Omega^2,
\end{equation}
where $n=0,1,2,\ldots$ and $\gamma$ is the adiabatic index.  
In the special case of purely vertical
perturbations, the inequality becomes an equality,
\begin{equation}
\omega^2=(n\gamma+1)\Omega^2.
\label{eq:vert_modes}
\end{equation}
The $n=0$ mode corresponds to a bulk vertical displacement of the disk
and subsequent vertical oscillation at the vertical epicyclic
frequency which, in the analysis of LP93 and in all analyses performed
in this paper, coincides with the orbital frequency.  In general, the
$n-th$ mode has $n$ vertical nodes (i.e., locations where the vertical
velocity perturbation vanishes), and is either even or odd depending
on whether $n$ is even or odd, respectively.

As discussed in \S~\ref{sec:mhd}, our simulations demonstrate the
effectiveness with which MHD turbulence excites these local acoustic
waves.

\subsection{Global oscillation modes of an accretion disk}

As discussed in the Introduction, several groups have studied the
global oscillation of black hole accretion disks using linear
perturbation theory, identifying three classes of normal mode
($g$-modes, $p$-modes and $c$-modes).  Trapped $g$-modes have received
particular attention as a possible source of HFQPO, although the other
families of modes may well be relevant.  Here we review some of the
basic results of these analyses, following the approach of Nowak \&
Wagoner (1991, 1992; hereafter NW91 and NW92).  The NW91 and NW92
analyses are not fully relativistic, instead employing a
pseudo-Newtonian potential.  Thus, these analyses can be readily
compared with our pseudo-Newtonian simulations.  We also note that
full general relativistic MHD simulations have typically yielded
results for slowly rotating black holes that are very similar to those
obtained with pseudo-Newtonian potentials (e.g., Gammie, McKinney, \&
T\'oth 2003; De~Villiers \& Hawley 2003).

NW91 and NW92 use a Lagrangian formalism (Friedman \& Schutz 1978) and
a WKBJ approximation to derive the linearized equations describing
perturbations of an inviscid hydrodynamic thin accretion accretion
disk about a pure Keplerian background state.  They also initially
examined the special case of purely radial oscillations and found the
standard dispersion relation of disk theory
(eqn.~\ref{eq:disk_dispersion}).  Given that the gravity-modified
$p$-modes described by this dispersion relation become evanescent when
$\omega^2<\kappa^2$, global $p$-modes can be trapped between the ISCO
(where $\kappa=0$) and the radius at which $\omega=\kappa$.  In
practice, however, the ``leaky'' nature of the ISCO would seem to make
the trapping of these modes ineffective.

The more general case, including perturbations that have vertical as
well as radial motions, yields more promising results.  NW92 examine
the linearized equations describing the behavior of the scalar potential
\begin{equation}
\delta u\equiv \delta P/\rho,
\end{equation}
where $\delta P$ is the Eulerian variation in the pressure.  They show
that the linearized equations are approximately separable into radial
and vertical equations, with the separation constant being a slowly
varying function of $r$, $\Upsilon(r)$.  The general dispersion
relation for these modes becomes
\begin{equation}
[\omega^2-\gamma \Upsilon(r)\Omega^2](\omega^2-\kappa^2)=\omega^2c_s^2k^2\; .
\end{equation}
Assuming that $\gamma \Upsilon\Omega^2>\kappa^2$, non-evanescent
solutions exist for $\omega^2>\gamma\Upsilon\Omega^2$ (predominantly
radial $p$-modes) or $\omega^2<\kappa^2$ (predominantly vertical
$g$-modes).  Through this analysis, NW92 identify a class of global
$g$-modes that are trapped between two evanescent regions, $r<r_-$ and
$r>r_+$, where $\kappa(r_\pm)=\omega$.  In other words, these modes
are trapped under the peak of the epicyclic frequency.  They
principally focus on the $m=0$ (axisymmetric) modes, and show that the
mode frequency is only {\it slightly} smaller than the maximum radial
epicyclic frequency $\kappa_{\rm max}$.  Radial harmonics of these
modes are very closely spaced.  Thus, the inner radius at which the
mode becomes evanescent is still a finite distance (and, in plausible
settings, several vertical scale heights) from the ISCO.  This raises
the interesting possibility of having appreciable power in such modes
without significant leakage across the ISCO.

A major issue, however, is the effect of the turbulent MHD background
state on these modes.  The diskoseismic mode frequencies are
comparable to the frequencies characterizing the expected MHD
turbulent fluctuations (which is very different to the situation in
the Sun, for example, where the observed helioseismic modes have
frequencies that are four orders of magnitude higher than the
turbulent turnover frequency).  Thus, an MHD turbulent disk is likely
to be a hostile environment for any diskoseismic modes.  Furthermore,
magnetic forces can lead to a rather gradual transition in flow
properties around the ISCO, potentially worsening the leakage of the
trapped g-modes.  Considering these mode destruction/suppression
mechanisms, it is reasonable to suppose that mode survival becomes
easier in thinner disks since (1) the ratio of the typical turbulent
cell size to the radial extent of the resonant cavity will decrease
with disk thickness and (2) there are suggestions that the transition
in flow properties around the ISCO is sharper in thinner disks
(Reynolds \& Fabian 2008; Shafee et al. 2008) thereby producing less
mode leakage.  This raises the possibility that a $g$-mode can be
sustained against (or even fed by; Nowak \& Wagoner 1993) the
turbulence in a sufficiently thin disk.

Previously published global MHD disk simulations (e.g., Hawley \&
Krolik 2001) have modeled flows as thin as $h/r\sim 0.1$ and have not
reported diskoseismic modes.  However, these authors did not conduct a
directed search for such modes and hence it is not possible to say
that the modes were really not present.  Careful examination of local
``shearing-box'' simulations have indeed failed to find trapped
$g$-modes associated with MHD turbulence (Arras, Blaes \& Turner
2006), but this issue has yet to be examined in a global thin-disk
setting.  Searching for and characterizing these trapped $g$-modes in
global thin-disk simulations will be a major theme of our paper.

\subsection{Parametric resonance}

From the point of view of explaining HFQPOs, the need for global
oscillation modes is diminished if some process does indeed select
special radii in the accretion disk.  As discussed in the
Introduction, the discovery of pairs of HFQPOs with small integer
ratios has prompted several groups to examine resonance
models.  In particular, we shall briefly review the parametric
resonance model of Abramowicz \& Klu\'zniak (Abramowicz \& Klu\'zniak
2001, 2003; Abramowicz et al. 2002, 2003).

We begin by considering an accretion disk in which the flow deviates
only slightly from Keplerian so that the position of a fluid element
(in spherical polar coordinates) is given by
\begin{equation}
r(t)=r_0+\delta r(t), \ \theta(t)=\frac{1}{2}\pi+\delta\theta(t),\ 
\phi(t)=\Omega t.
\end{equation}
We have specialized to the case of axisymmetric perturbations.  By
expanding the resulting equations of motion to third order (and
wrapping up the non-gravitational forces into two unspecified force
functions), one finds a Mathieu-type equation of motion
\begin{equation}
\delta\theta_{,\,tt}+\Omega^2[1+a\cos(\kappa t)]\delta\theta +
\lambda\,\delta\theta_{,\,t}=0,
\label{eq:mathieu}
\end{equation}
where $a$ is a small constant that describes the coupling between the
vertical and radial perturbations, and $\lambda$ is a (small) damping
constant.  Here, we have specialized the equations of Abramowicz et
al. (2003; hereafter A2003) to the case where the vertical epicyclic
frequency is the same as the orbital frequency.  This is valid for a
non-spinning black hole and, in particular, the pseudo-Newtonian
potential that we use for the simulations in this paper.

One expects a system described by eqn.~\ref{eq:mathieu} to exhibit a
parametric resonance instability when $\kappa=2\Omega/n$, where $n$ is
a non-zero positive integer.  Given that black hole potentials always
have $\kappa<\Omega$, the smallest value of $n$ for which the
resonance condition is obeyed is $n=3$, i.e., $\kappa=2\Omega/3$.
This is expected to be the strongest of the set of resonances.  It is
interesting that the fundamental ``test-particle'' frequencies at this
resonant radius have 3:2 frequency ratio in agreement with
observations of HFQPO pairs.  For the Paczynski-Wiita pseudo-Newtonian
potential we use in our simulations (Paczynski \& Wiita 1980, hereafter PW),
\begin{equation}
\Phi=-\frac{GM}{r-2r_g},\hspace{1cm}r_g\equiv \frac{GM}{c^2},
\label{eq:pw_pot}
\end{equation}
we have
\begin{equation}
\Omega=\frac{1}{r-2r_g}\sqrt{\frac{GM}{r}}
\end{equation}
and,
\begin{equation}
\kappa=\sqrt{\frac{GM(r-6r_g)}{r(r-2r_g)^3}}.
\end{equation}
Hence, the 3:2 resonance occurs at $r=9.2r_g$. It must be noted,
however, that a full integration of a toy-problem by A2003 reveals
that higher-order effects shift the location of the resonance, making
the ratio of the epicyclic frequencies extremely sensitive to the
strength of the coupling between the $r$ and $\theta$ perturbations.
While A2003 suggest that this sensitivity is a strength of the model,
allowing application to HFQPO pairs in neutron star systems which do
not have simple integer ratios, it inevitably diminishes the power of
this model to explain black hole systems.  In addition, there is
currently no physical model of the coupling between the radial
and vertical motions.  The simulations described in this paper allow
us to assess whether magnetic forces couple these motions in such a
way as to drive a parametric resonance instability.

\section{Hydrodynamic disks}

\subsection{Initial comments}
\label{sec:initial_hydro}

In the remainder of this paper, we construct and analyze numerical
simulations of thin disks in order to examine their variability
properties, focusing on the presence of local and global modes, as
well as parametric resonances.  Of course, real accretion disks are
believed to require at least an MHD-level description in order to
capture the MRI-driven turbulence that transports angular momentum and
drives accretion.  MHD simulations are addressed in the next section.
However, it is useful to begin with a discussion of ideal hydrodynamic
models in order to help isolate the various physical effects present
in these complex systems.  That will be the focus of this section.

In order to allow us to perform a set of simulations with
modest-to-high resolution, we begin with 2-d (axisymmetric)
hydrodynamic simulations.  We expect (and indeed show) that these
axisymmetric models are well suited for studying the fundamental $m=0$
g-mode.  However, once we move to MHD, Cowling's anti-dynamo theorem
(Cowling 1957) leads to qualitatively different behavior in
axisymmetric compared with full 3-d simulations.  Hence, all of the
MHD models that we shall describe are performed in 3-dimensions.
Since we will be directly comparing results (e.g., $g$-mode
amplitudes) between the hydrodynamic and MHD simulations, we also need
to perform a ``bridging'' 3-d hydrodynamic simulation.

\subsection{Basic simulation set-up}
\label{sec:hydro_setup}

To make the problem tractable despite the severe resolution
requirements imposed by the geometrical thinness of the accretion
disk, we choose to focus on only the most essential aspects of the
physics.  From the discussion in \S~2, it is clear that the essential
aspect of the relativistic potential that must be captured is the
nature of the radial epicyclic frequency (e.g., that it goes to zero
at some finite radius and hence produces an ISCO at that radius).  In
this sense, the PW potential (eqn.~\ref{eq:pw_pot}) is a good
approximation for the gravitational field of a non-rotating black
hole; its ISCO (at $6r_g$) and marginally bound orbit (at $4r_g$) are
both at the same radius as in the Schwarzschild geometry. We also
simplify the simulations by neglecting all radiation processes
(radiative heating, radiative cooling, radiative transfer, and the
dynamical effects of radiation pressure).  In place of a full energy
equation, the gas is given an adiabatic equation of state with
$\gamma=5/3$.

\begin{figure*}
\hbox{
\hspace{-1cm}
\psfig{figure=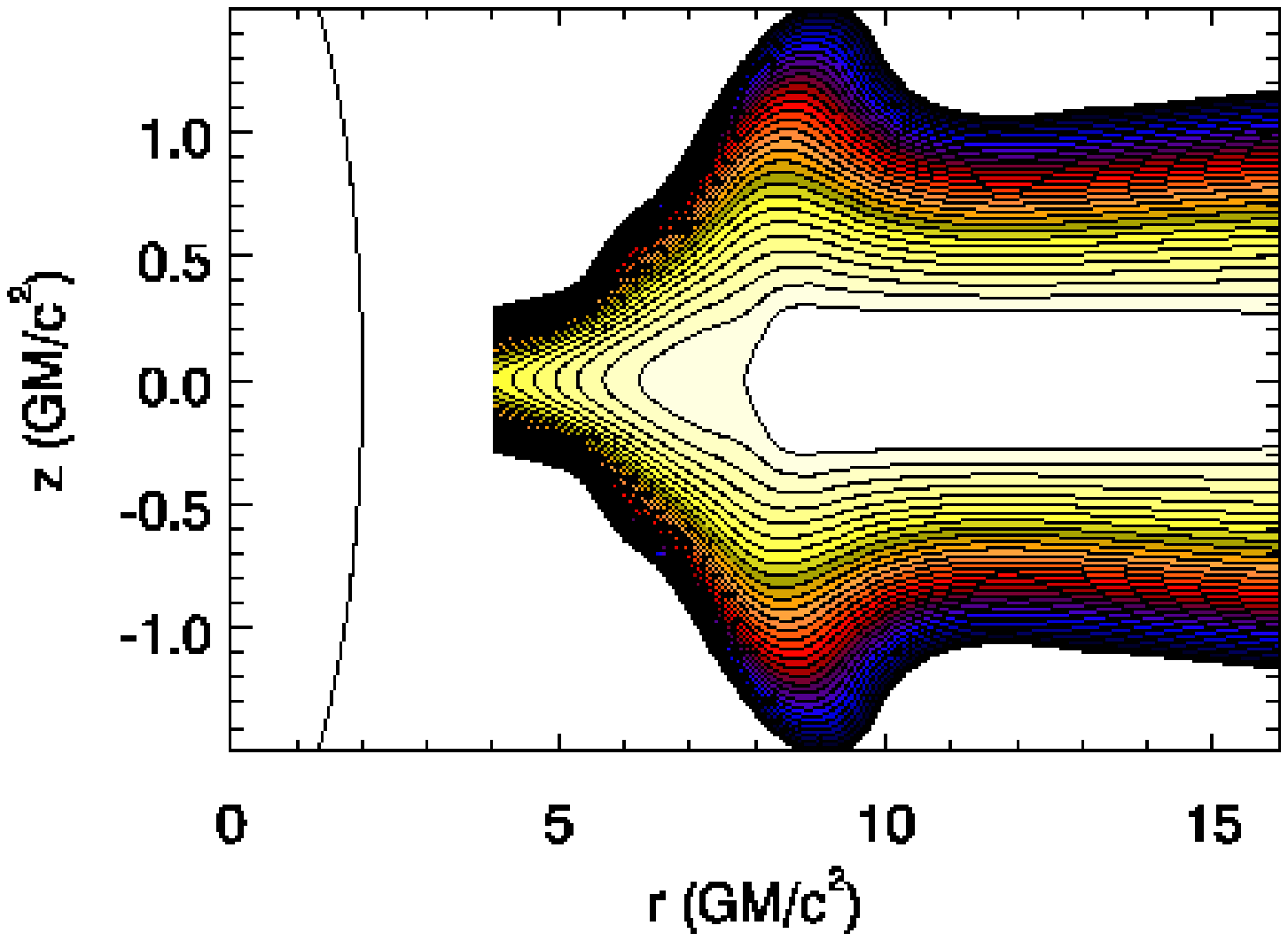,width=0.55\textwidth}
\hspace{-1cm}
\psfig{figure=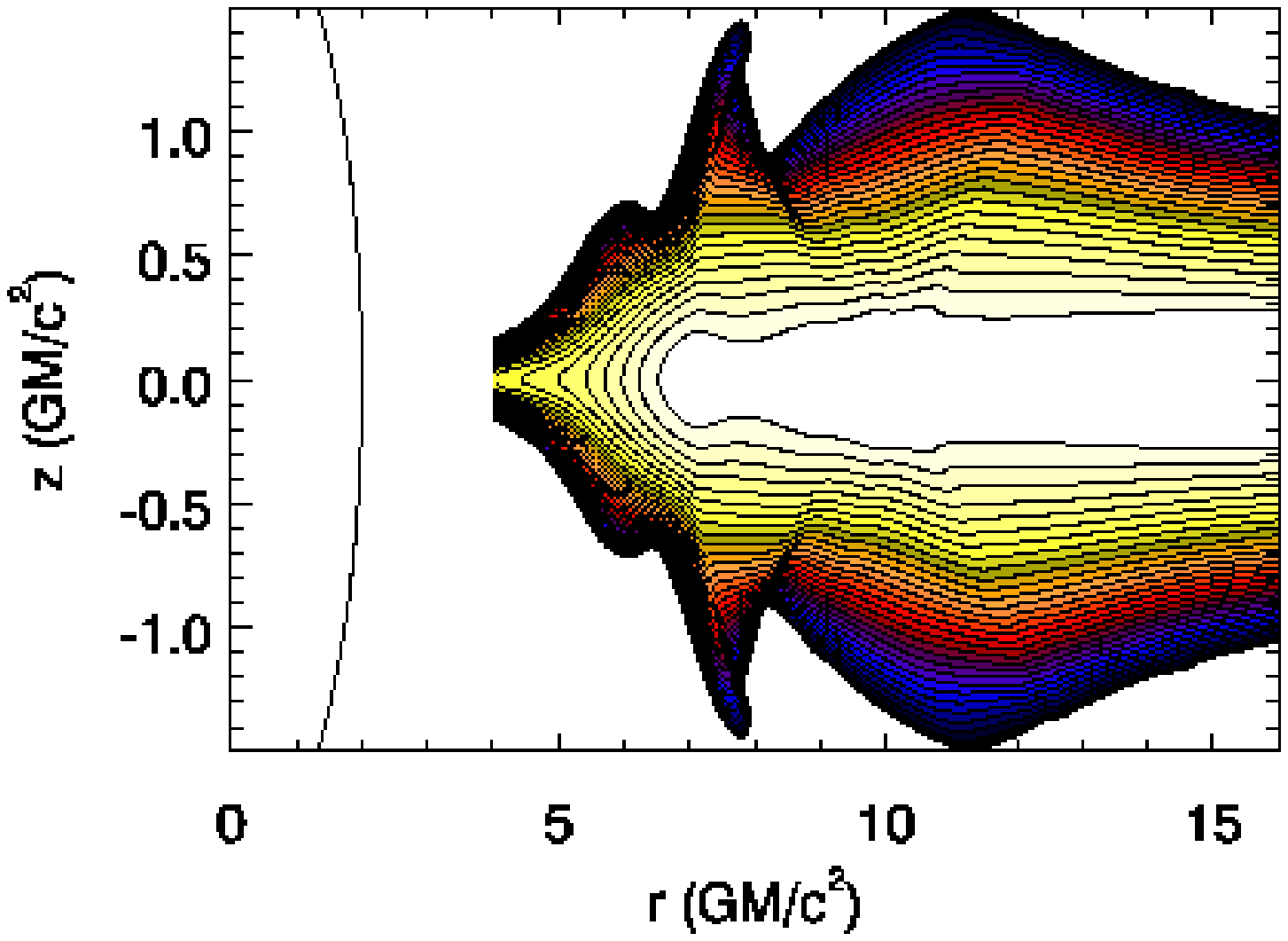,width=0.55\textwidth}
}
\hbox{
\hspace{-1cm}
\psfig{figure=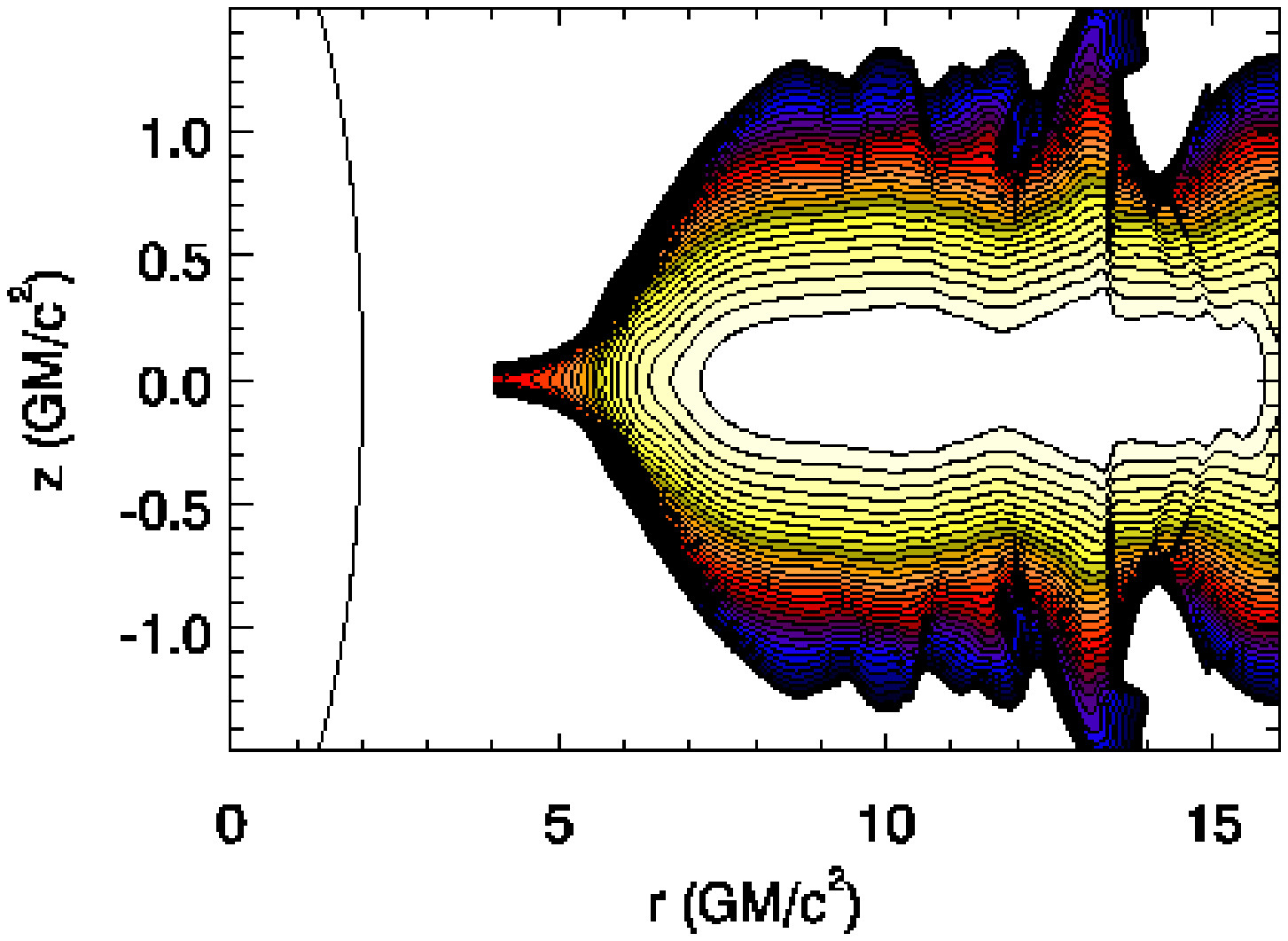,width=0.55\textwidth}
\hspace{-1cm}
\psfig{figure=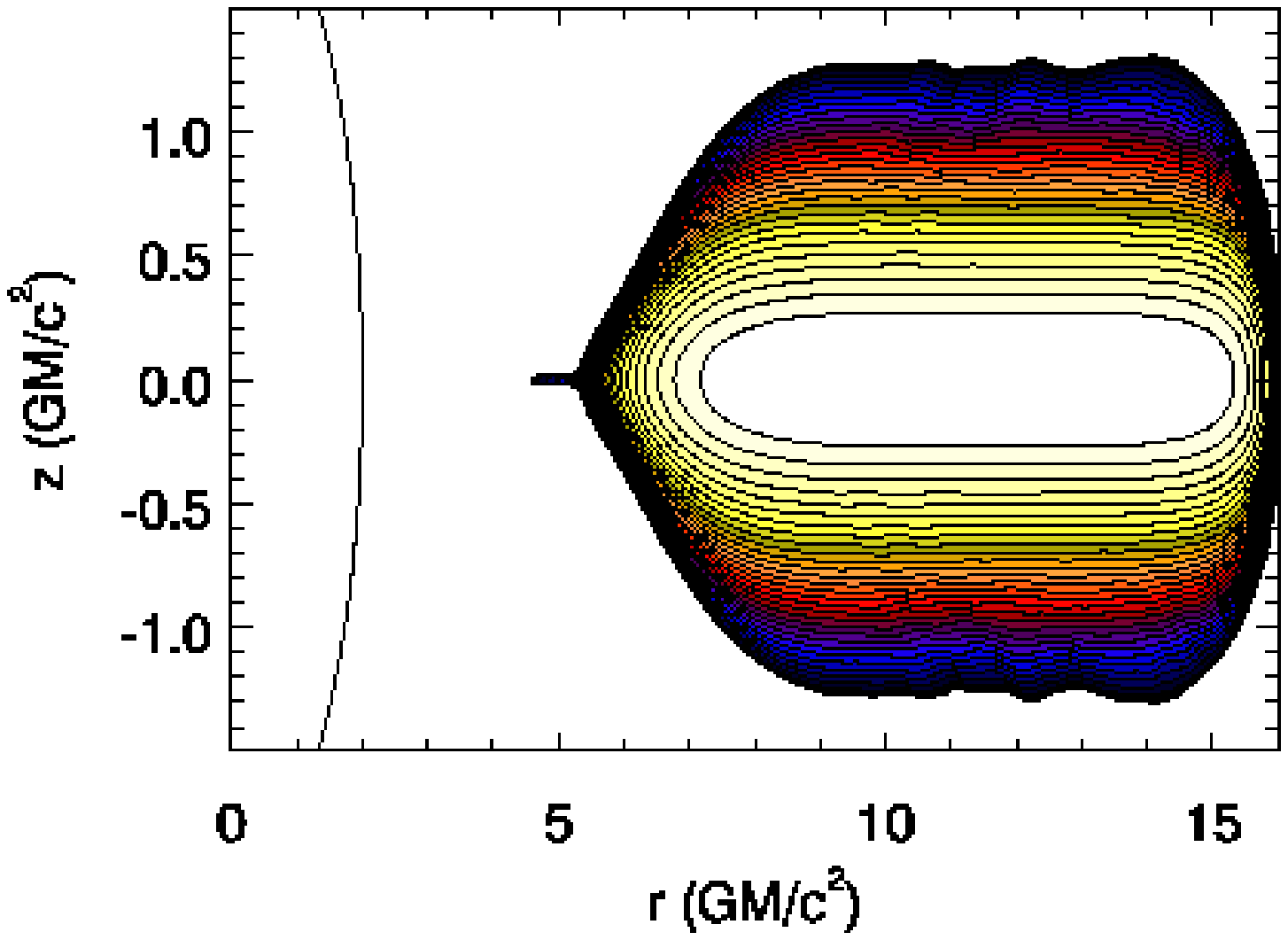,width=0.55\textwidth}
}
\caption{Snapshots of evolution of the canonical axisymmetric
hydrodynamic simulation (HD2d\_1) at $t=0.5T_{\rm isco}$ (top-left),
$t=1.0T_{\rm isco}$ (top-right), $t=5T_{\rm isco}$ (bottom-left) and
$t=50T_{\rm isco}$ (bottom-right), where $T_{\rm isco}$ is the orbital
period at the ISCO.  Both the color table and contours show the
logarithmic density structure of the disk cross-section, with 10
contours per decade of density.  A range of densities spanning three
orders of magnitude are shown.  The curved line to the left of each
frame represents the event horizon.}
\label{fig:hydro_evol}
\end{figure*}

\begin{table*}[t]
\begin{center}
{\small
\begin{tabular}{cccccccc}\hline
Run     &   dim   & $h_2/r_{\rm isco}$ & r-domain ($r_g$)& z-domain ($r_g$)     & $\phi$-domain & $n_r\times n_z (\times n_\phi)$ & $T_{\rm stop}$ ($r_g/c$) \\
(1)     &    (2)  &   (3)              &   (4)           &    (5)               &     (6)       &      (7)                        &  (8)           \\\hline
HD2d\_1     &  2  & 0.05$^*$  & (4,16)   & (-1.5,1.5)    &    --         & $256\times 128$                 & 12320 \\
HD2d\_1hr   &  2  & 0.05$^*$  & (4,28)   & (-1.5,1.5)    &    --         & $1024\times 256$                & 12320 \\
HD2d\_2     &  2  & 0.025$^*$ & (4,16)   & (-0.75,0.75)  &    --         & $512\times 128$                 & 12320 \\
HD2d\_3     &  2  & 0.1$^*$   & (4,16)   & (-3,3)        &    --         & $256\times 128$                 & 12320 \\
HD3d\_1     &  3  & 0.05$^*$  & (4,16)   & (-1.5,1.5)    & $(0,\pi/6)$   & $240\times 128\times 32$        & 12320 \\\hline
MHD\_1 &       3  & 0.05  & (4,16)   & (-3,3) &    $(0,\phi/6)$  & $240\times 256\times 32$     & 38800  \\
MHD\_2 &       3  & 0.05  & (4,16)   & (-1.5,1.5) &    $(0,\phi/6)$  & $240\times 128\times 32$     & 38800  \\
MHD\_2hr &     3  & 0.05  & (4,16)   & (-1.5,1.5) &    $(0,\phi/3)$  & $480\times 256\times 64$     & 5236 \\
MHD\_3 &       3  & 0.05  & (4,28)   & (-1.5,1.5) &    $(0,\phi/6)$  & $960\times 128\times 32$     & 11400 \\\hline
\end{tabular}
\caption{Summary of the hydrodynamic (HD) and MHD simulations
discussed in this paper.  Column (1) gives the designation of the
simulation.  Column (2) lists the dimensionality of the simulation.
Column (3) gives the fractional disk thickness at the ISCO; the
asterisk ($*$) denotes that the simulation was started with an initial
vertical structure that is slightly out of equilibrium in order to
seed subsequent oscillation modes (as described in the text).  Columns
(4), (5) and (6) list the $r$-, $z$- and $\phi$- domains of the
simulation box.  Column (7) gives the number of computational zones
within the domain.  Column (8) lists the run time of the simulated
disk.}}
\end{center}
\end{table*}

All simulations are performed in cylindrical polar coordinates
$(r,z,\phi)$.  The initial condition consists of a disk with a
constant mid-plane density ($\rho_0=1$) for $r>r_{\rm isco}$.  The
initial density scale-height of the disk is assumed to be constant
with radius, implying a sound speed which falls off with radius as
approximately $r^{-3/2}$.  There are two motivations for choosing to
model a ``constant-h'' disk; (1) such a choice is well suited to the
cylindrical symmetry of our coordinate grid and (2) according to the
standard model of Shakura \& Sunyaev (1973), the radiation-pressure
dominated disks of real accreting black holes are likely to maintain
an approximately constant scale-height in their innermost regions.
In more detail, the vertical structure of the disk is given by
\begin{eqnarray}
\rho(r,z)&=&\rho_0\,\exp\left(-\frac{z^2}{2h_1^2}\right),\\
p(r,z)&=&\frac{GMh_2^2}{(R-2r_g)^2R}\,\rho(r,z),
\end{eqnarray}
where $r$ is the cylindrical radius, $z$ is the vertical height above
the disk midplane and $R=\sqrt{r^2+z^2}$. This corresponds to an
isothermal atmosphere which, when $h_1=h_2$, is in vertical
hydrostatic equilibrium in the PW potential.  In order to give the
disk an initial vertical kick, we set $h_1=1.2h_2$ (the values of
$h_2$ for all of our runs are detailed in Table~1).  Thus, the initial
disk temperature is $\sim 20\%$ too cold for the density and pressure
run, leading to a vertical collapse and bounce of the disk.  The
initial density is set to zero for $r<r_{\rm isco}$.  The initial
velocity field is everywhere set to
\begin{equation}
v_\phi=r\Omega=\frac{\sqrt{GMr}}{r-2r_g},\hspace{1cm}v_r=v_z=0,
\end{equation}
corresponding to rotation on cylinders and pure Keplerian motion for
material on the mid-plane.  We impose zero-gradient outflow boundary
conditions on both the radial and vertical boundaries of the
simulation, i.e., the fluid quantities in the ghost zones are set to
the values of the neighboring active zone, and a ``diode'' condition
is imposed on the component of the velocity perpendicular to the
boundary which allows outflow but disallows inflow.

Table~1 details our hydrodynamic simulations.  We perform four
simulations in which strict axisymmetry is imposed at all times
($\partial/\partial \phi=0$).  These axisymmetric simulations were
performed using the serial ZEUS-3D MHD code (Stone \& Norman 1992a,b)
in its pure hydrodynamic axisymmetric mode.  In our canonical
axisymmetric run (HD2d\_1), we set $h_2=0.3r_g$ (corresponding to
$h_2/r=0.05$ at the ISCO).  To model the dynamics of the disk in a
robust manner, our vertical domain must cover many scale-heights; in
our canonical run, the vertical domain is $z\in (-5h_2,+5h_2)$.  We
place 128 uniformly spaced grid cells in this vertical direction,
giving 13 cells per scale-height.  This allows us to resolve
hydrodynamic waves with wavelengths of $\sim 0.5h_2$ or greater.  The
vertical resolution requirements force us to consider a limited radial
domain.  In the canonical simulation, the radial domain extends from
$4r_g$ (i.e., well within the plunge region) to $16r_g$ and should
correspond to the range of radii where trapped diskoseismic modes or
A2003-type resonances occur.  Tolerating a grid-cell aspect ratio of
$\sim 2$, we place 256 uniformly spaced radial cells in this radial
domain.  The simulation was evolved for a time $200T_{\rm isco}$ where
\begin{equation}
T_{\rm isco}\approx 61.6\,GM/c^3,
\end{equation} 
is the orbital period at the ISCO.

In order to test the robustness of the results discussed below to
resolution and the limited radial domain, we performed a second
simulation (HD2d\_1hr) in which the spatial resolution was doubled
(i.e., the size of each voxel was halved in both the radial and
vertical dimensions) and the radial domain extended out to $28r_g$.
We also performed two additional axisymmetric simulations employing
the same set-up as the canonical simulation but with disks that are
half (HD2d\_2) and twice (HD2d\_3) the thickness (including
appropriate modifications to the vertical domain and resolution; see
Table~1).

As discussed above, we perform a 3-dimensional hydrodynamic simulation
in order to aid the later interpretation of the 3-d MHD simulations.
The set-up of our 3-d run (HD3d\_1) is identical to that for the
canonical 2-d run except that the computational domain has a
$\phi$-dimension.  To reduce computational expense while capturing the
essential physics, we simulate only a $\Delta\phi=30^\circ$ wedge of
the disk using 32 uniformly spaced grid cells, imposing periodic
boundary conditions on the $\phi$-boundaries.  This 3-dimensional
simulation was performed using an MPI-parallelized version of ZEUS
kindly supplied to us by Eve Ostriker (and similar to the ZEUS-MP code
of Vernaleo \& Reynolds 2006).

\subsection{Axisymmetric hydrodynamic models and the recovery of trapped 
$g$-modes}
\label{sec:axisym}

We now discuss the evolution of the axisymmetric hydrodynamic
simulations, beginning with the canonical simulation.  Starting from
the initial condition, the disk undergoes dynamical timescale
variability because of pressure gradients which push matter inside of
the ISCO.  The strong transients close to the ISCO launch
outward-radially directed waves into the disk which break rapidly to
become rolls.  This behavior can be seen in Fig.~\ref{fig:hydro_evol}.
These high amplitude transients are short lived, however, lasting only
$\sim 10T_{\rm isco}$.  At subsequent times, the disk settles into a
stationary state apart from small amplitude (and decaying) internal
oscillations and a very weak accretion stream driven by the numerical
viscosity.

\begin{figure}[t]
\centerline{
\psfig{figure=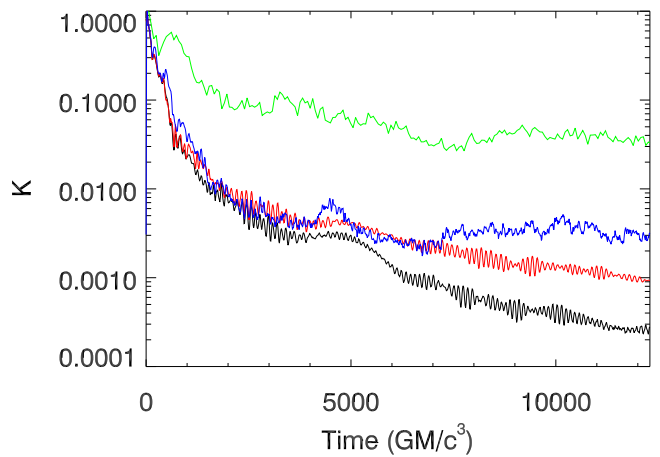,width=0.55\textwidth}
}
\caption{Change of the quantity $K=\int_{\cal D} \rho v_z^2\,dV$ with
time.  The integration domain ${\cal D}$ is the annulus $r\in (7,14)$.  The
bottom (black) line is for the canonical-2d run (HD2d\_1),
lower-middle (red) line is for the high-resolution 2-d run
(HD2d\_1hr), the upper-middle (blue) line is for the canonical 3-d run
(HD3d\_1), and the top (green) line is for the canonical MHD run
(MHD\_1).  The distinct ``ringing'' seen in runs HD2d\_1 and HD2d\_1hr
corresponds to the axisymmetric g-mode (see Section~\ref{sec:axisym}).
In run HD3d\_1, non-axisymmetric aperiodic structures mask the
underlying g-mode (see Section~\ref{sec:3dhydro}).  The fact that
$K(t)$ for the run MHD\_1 lies an order of magnitude above that for
HD3d\_1 demonstrates that the MRI-driven turbulence completely
overwhelms the hydrodynamic disturbances seen in the analogous 3-d
hydrodynamic simulation (see Section~\ref{sec:basicmhdevol}).}
\label{fig:hydro_decay}
\end{figure}

In order to study the decay of the hydrodynamic fluctuations in a more
quantitative manner, we start by computing the quantity,
\begin{equation}
K=\int_{\cal D} \rho v_z^2\,dV.
\label{eq:decay}
\end{equation}
This quantity is particularly well suited to diagnose vertical
oscillations of the disk, and will vanish once the hydrodynamic
configuration has established a stationary state.  In order to
diagnose the state of the body of the disk (i.e., side-stepping issues
of the outer radial boundary or the plunge region), we choose to
compute this integral over a restricted domain ${\cal D}$ consisting
of the annulus $r\in (7r_g,14r_g)$.  The time-dependence of
$K/\max(K)$ for the canonical axisymmetric simulation (HD2d\_1) and its
high-resolution counterpart (HD2d\_1hr) is shown in
Fig.~\ref{fig:hydro_decay}.  The rapid initial decline of $K(t)$ is
very similar for these two simulations and corresponds to the strong
transients described above.  At long times (after about $t\sim 4\times
10^3\,GM/c^3\approx 70\,T_{\rm isco}$) the behavior of these
simulations begins to deviate.  HD2d\_1 continues to decay in an
approximately exponential manner $K(t)\propto e^{-t/\tau_0}$, where
$\tau_0\approx 4\times 10^3\,GM/c^3$.  Superposed on this decay is a
distinct oscillation.  This corresponds to (twice) the frequency of
the trapped $g$-mode that we shall discuss below.  The high-resolution
version of this simulation HD2d\_1hr shows very similar behavior
except that the exponential decay time is longer, $\tau_0\approx
6\times 10^3\,GM/c^3$.  This suggests that the decay of these small
perturbations is due to numerical dissipation which scales
approximately as the square root of the size of the simulation grid
cells.

\begin{figure}[b]
\centerline{
\psfig{figure=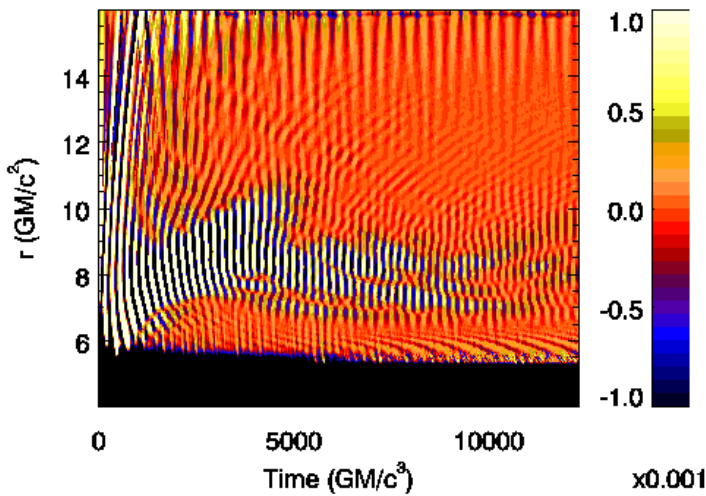,width=0.55\textwidth}
}
\caption{Radial component of velocity $v_r$ on the midplane of the
disk as a function of radius and time for the the canonical
axisymmetric hydrodynamic simulation (HD2d\_1).  The linear color
table extends from radial velocities of $v_r=-0.001c$ (black) to
$v_r=+0.001c$ (white).  Note that, outside of the ISCO, the average
value is $|v_r|\ll 0.001c$ so that this diagram essentially plots the
fluctuation of $v_r$ from its mean value.}
\label{fig:vr_rt}
\end{figure}

We now study the spatio-temporal variability of the disk and, in
particular, seek the diskoseismic modes predicted by linear theory.
Figure~\ref{fig:vr_rt} shows the mid-plane value of $v_r$ on the
$(r,t)$-plane, i.e., the value of the function $v_r(r,z=0;t)$, for run
HD2d\_1.  Note that, outside of the ISCO, the average value is
$\langle v_r\rangle\ll 0.001c$ so that this Figure essentially plots the
fluctuation of $v_r$ from its mean value.  At early times, we see
strong wave-like disturbances which are generated in the inner parts
of the disk (at $r\approx 8r_g$) and propagate to both large and small
radii.  Although the outer radial boundary condition is zero-gradient
outflow, impedance mismatching results in some reflection of these
initial strong waves.  After these initial transients have died out,
it can be seen that the highest amplitude fluctuations are limited to
a rather narrow range of radii in the approximate range $r=7-9r_g$.
The fact that these perturbations are essentially vertical on the
$(t,r)$-plane indicates that they are coherent across this radial
range, as would be expected if we are seeing a global mode.

\begin{figure*}
\hbox{
\psfig{figure=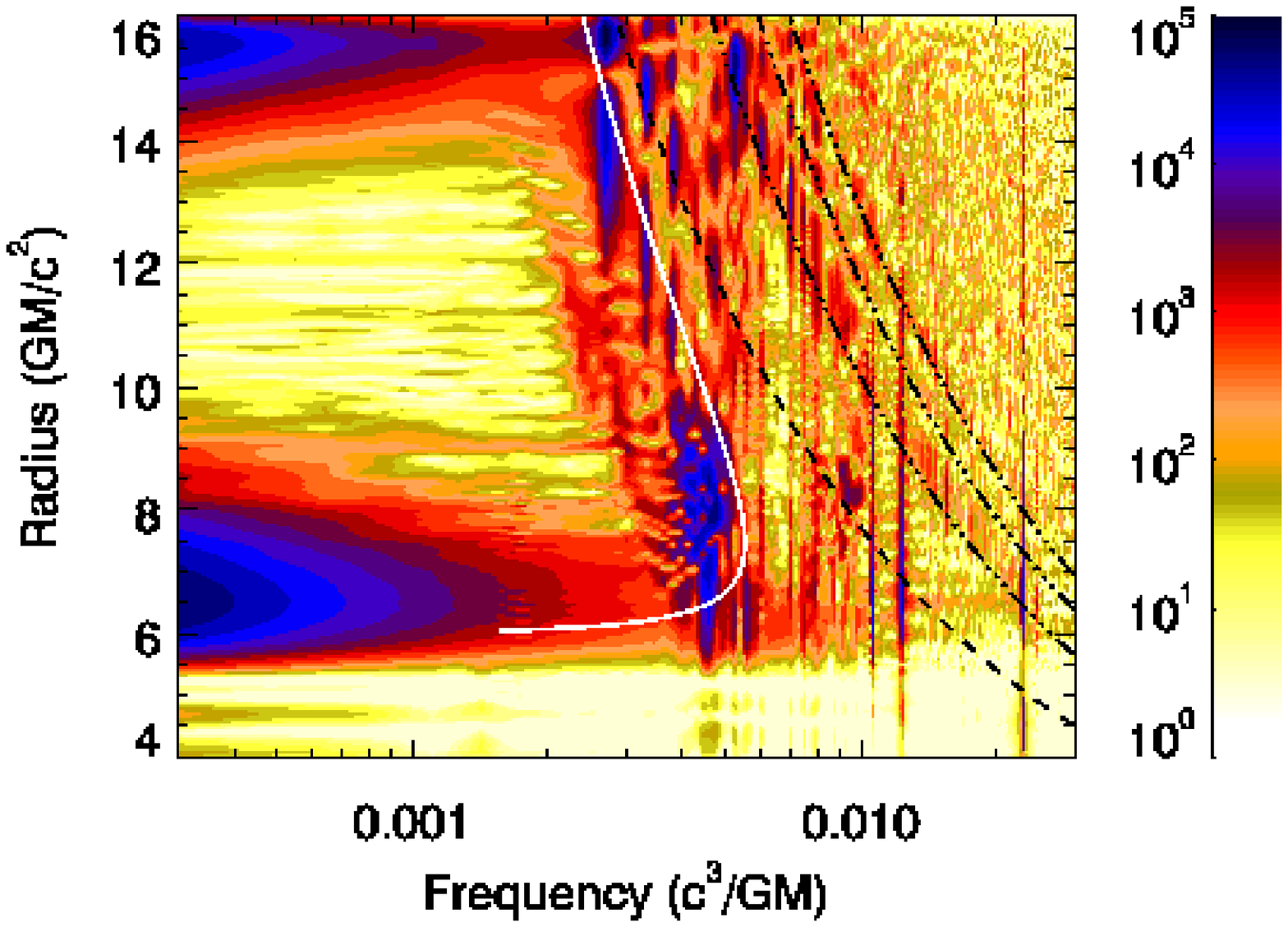,width=0.55\textwidth}
\psfig{figure=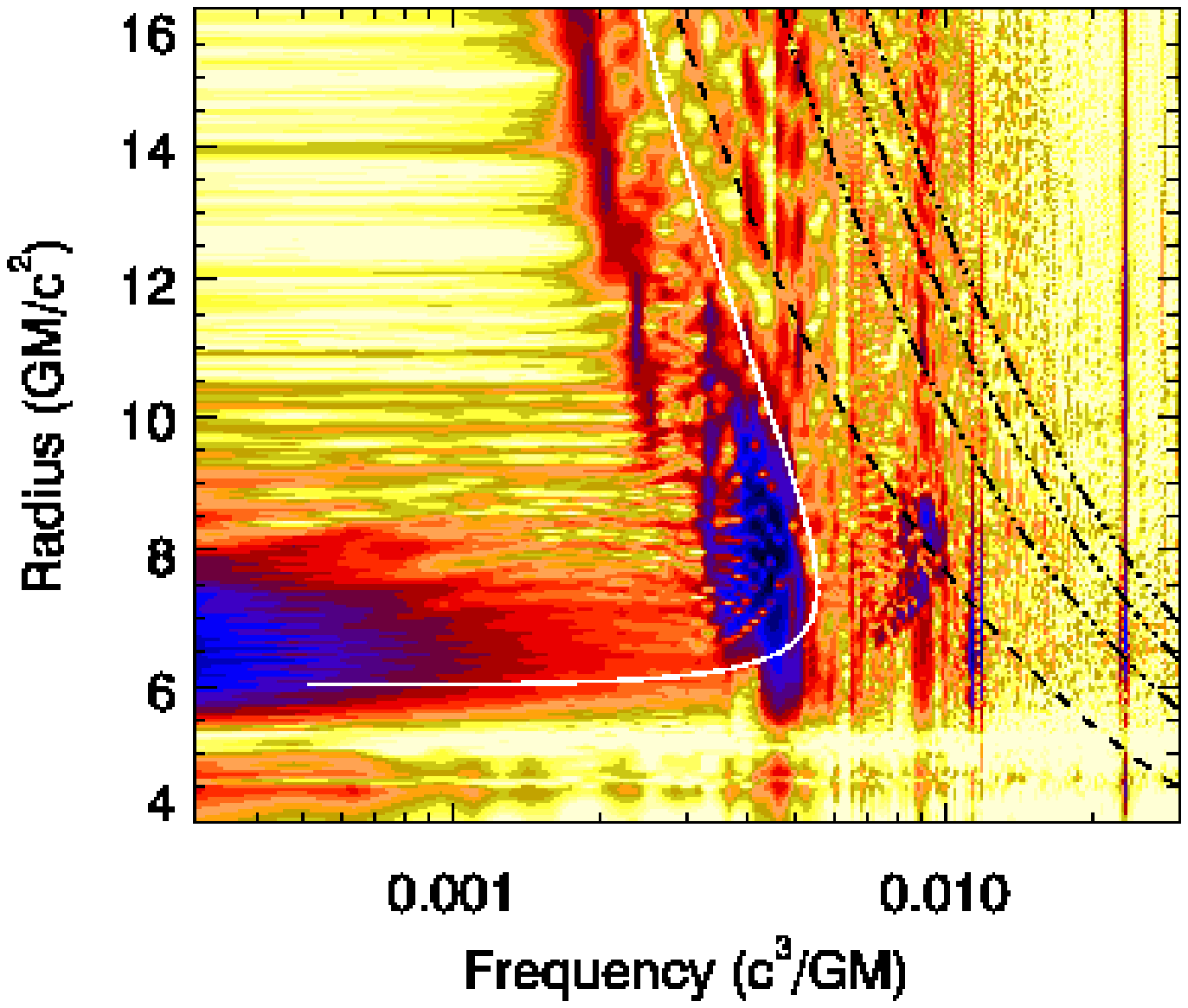,width=0.55\textwidth}
}
\caption{Midplane PSD for pressure fluctuations in run HD2d\_1 (left
panel) and its high-resolution counterpart HD2d\_1hr (right panel).
Note that we show the radial range $r\in (4,16)$ which is coincident
with the full computational domain of HD2d\_1 but only the inner half
of the domain for HD2d\_1hr (whose full radial domain extends from
$4r_g$ to $28r_g$). Also shown are the radial epicyclic frequency
(solid white line), orbital frequency (dashed) and the $n=1,2,3$ pure
vertical p-modes (from left to right dot-dashed lines).  The absolute
scaling of the PSD, as indicated by the color-bar, is arbitrary.}
\label{fig:hd_psd}
\vspace{0.5cm}
\end{figure*}

This temporal variability can be explored in more detail using the
power spectral density (PSD), defined as $P(\nu)=\alpha |\tilde
f(\nu)|^2$, where $\alpha$ is some normalization constant and $\tilde
f(\nu)$ is the Fourier transform of the time-sequence $f(t)$ under
consideration,
\begin{equation}
\tilde f(\nu)=\int f(t)e^{-2\pi i\nu t}\,dt.
\end{equation}
Note that, for ease of interpretation, most of the PSDs presented in
this paper will be in terms of the usual frequency, $\nu$, rather than
angular frequency, $\omega$.  Figure~\ref{fig:hd_psd} shows the PSD of
the mid-plane pressure as a function of $r$ and frequency for HD2d\_1
and HD2d\_1hr.  This is computed using approximately the final half
($\Delta t=102.4T_{\rm isco}\approx 6308\,GM/c^3$) of the simulation
in order to avoid the initial strong transients.  These PSDs differ in
the range $r=12-16r_g$; run HD2d\_1hr has significantly less low
frequency noise than run HD2d\_1 in this radial range.  We attribute
this to effects related to the outer radial boundary which is at
$r=16r_g$ in run HD2d\_1 but substantially further out ($r=28r_g$) in
run HD2d\_1hr.

\begin{figure*}
\hbox{
\hspace{-1cm}
\psfig{figure=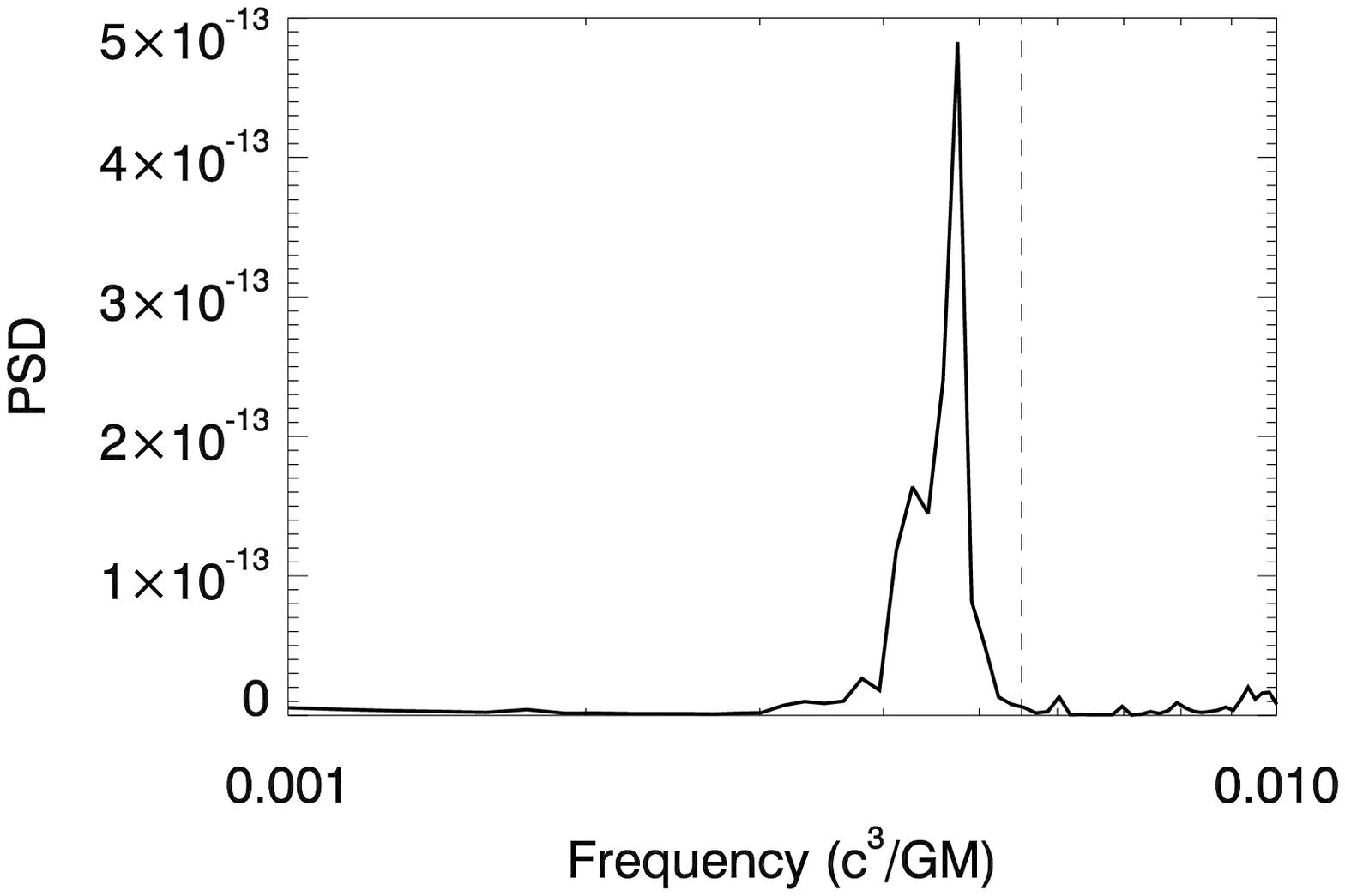,width=0.55\textwidth}
\hspace{-1cm}
\psfig{figure=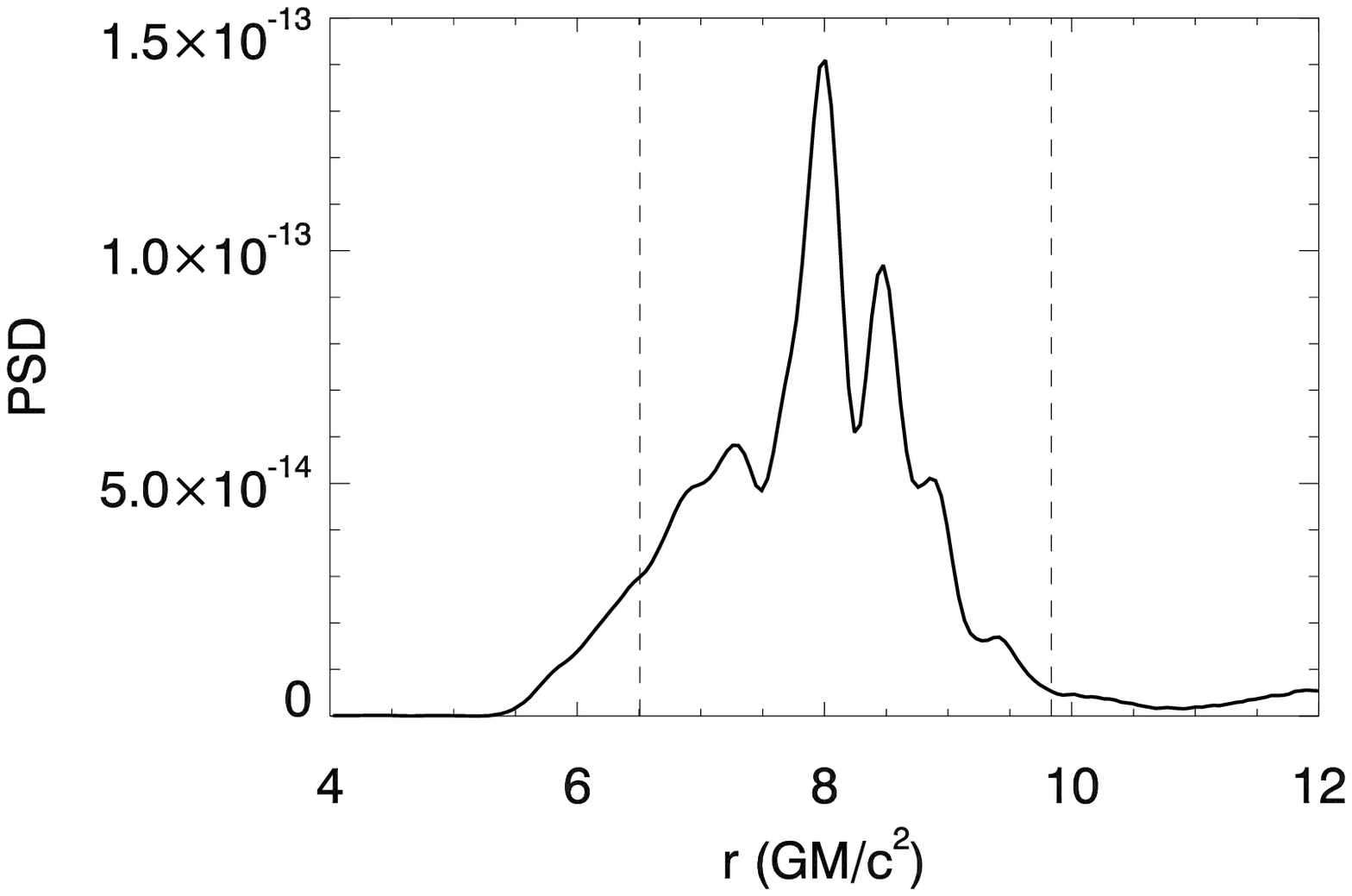,width=0.55\textwidth}
}
\caption{{\it Left panel} : Midplane pressure PSD for run HD2d\_1
summed up over a range of radii $\Delta r=0.5r_g$ centered on
$r=8r_g$.  The dashed line shows the maximum radial epicyclic frequency.
{\it Right panel }: Integral of the midplane pressure PSD of those
frequency bins that exceed $5\times 10^{-14}$ in the left panel, as a
function of radius. Vertical dashed lines show the range of radii for
which the mean frequency of this peak is less than the radial
epicyclic frequency.}
\vspace{0.5cm}
\end{figure*}

\begin{figure*}
\hbox{
\hspace{-1cm}
\psfig{figure=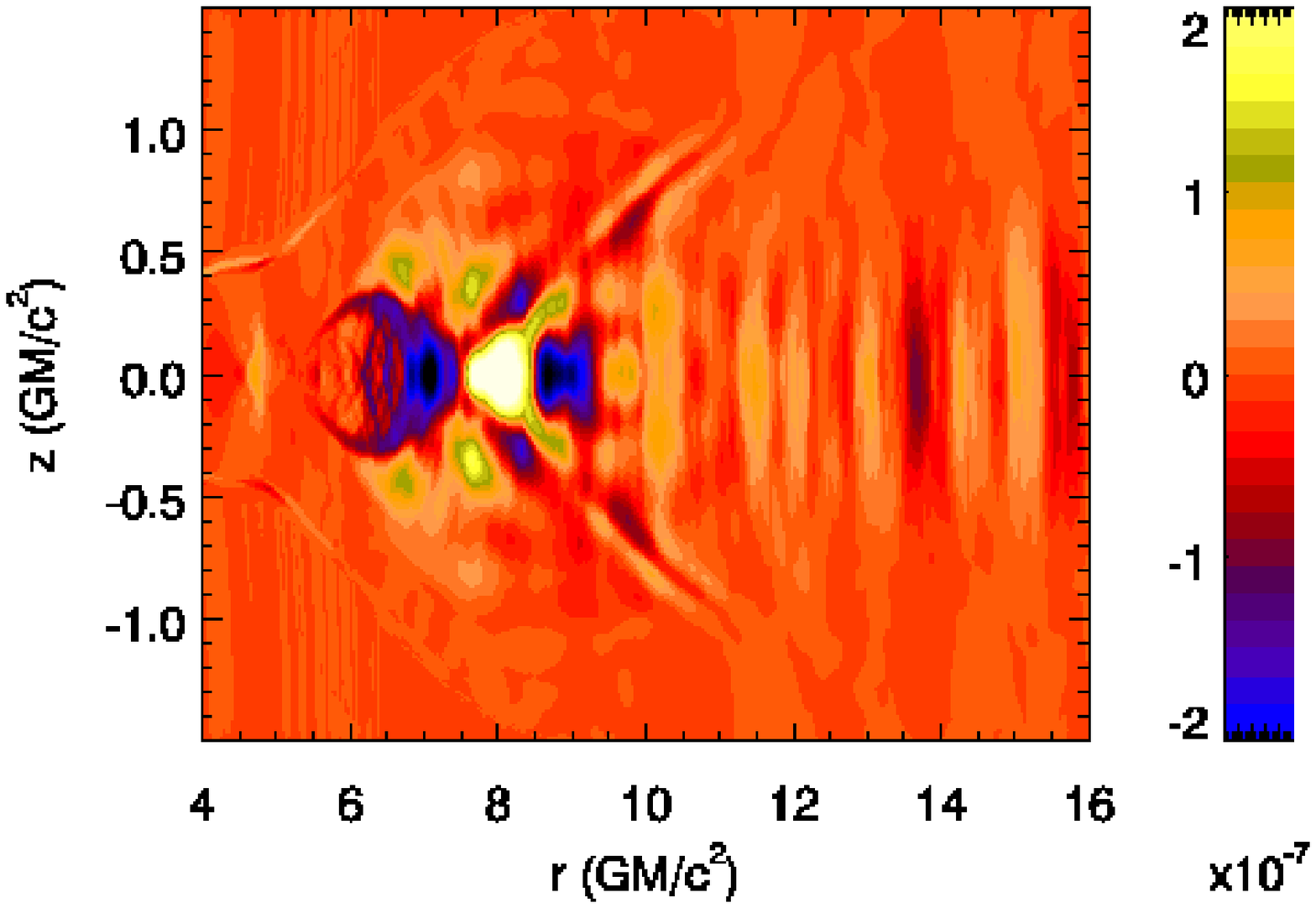,width=0.55\textwidth}
\psfig{figure=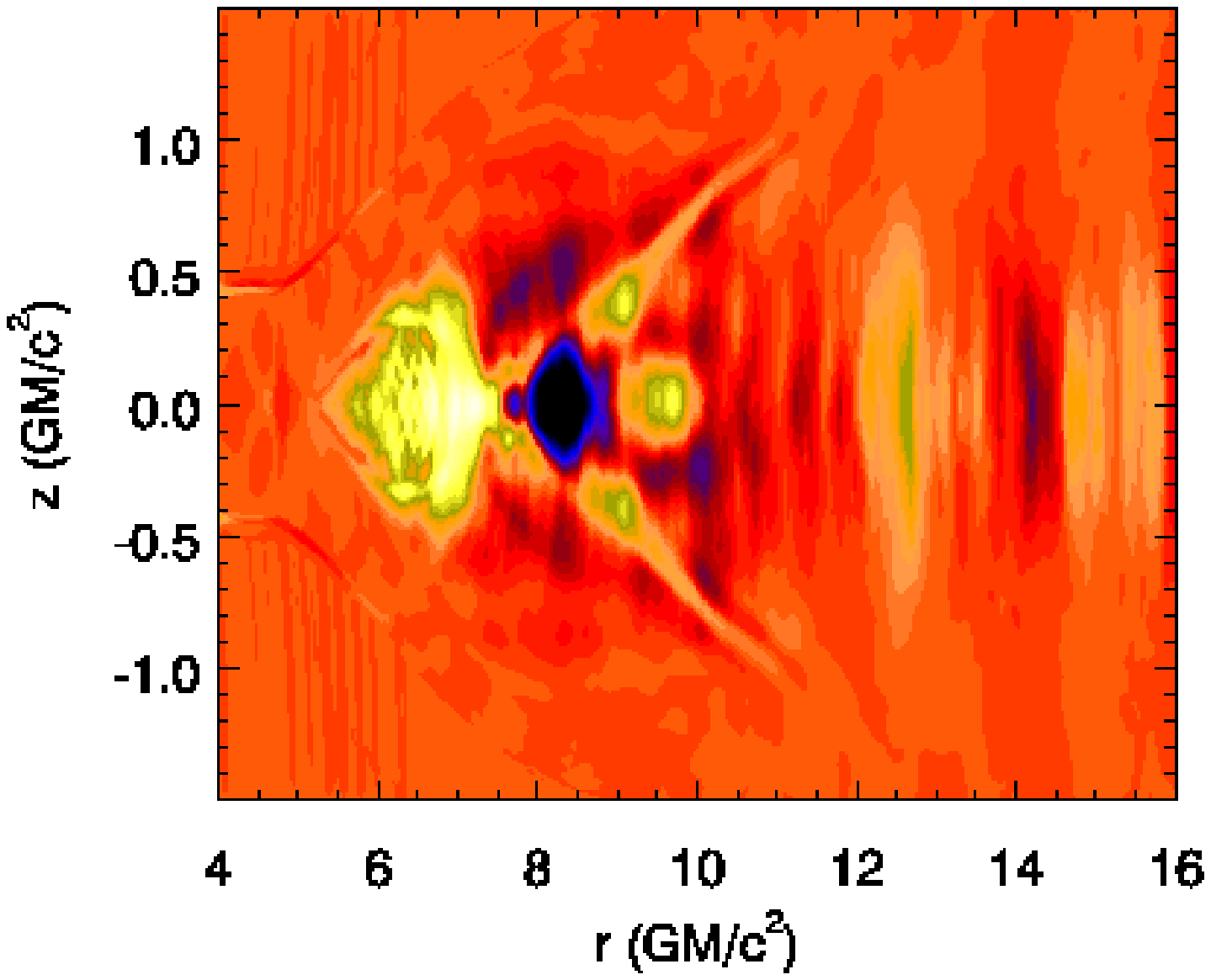,width=0.55\textwidth}
}
\hbox{
\hspace{-1cm}
\psfig{figure=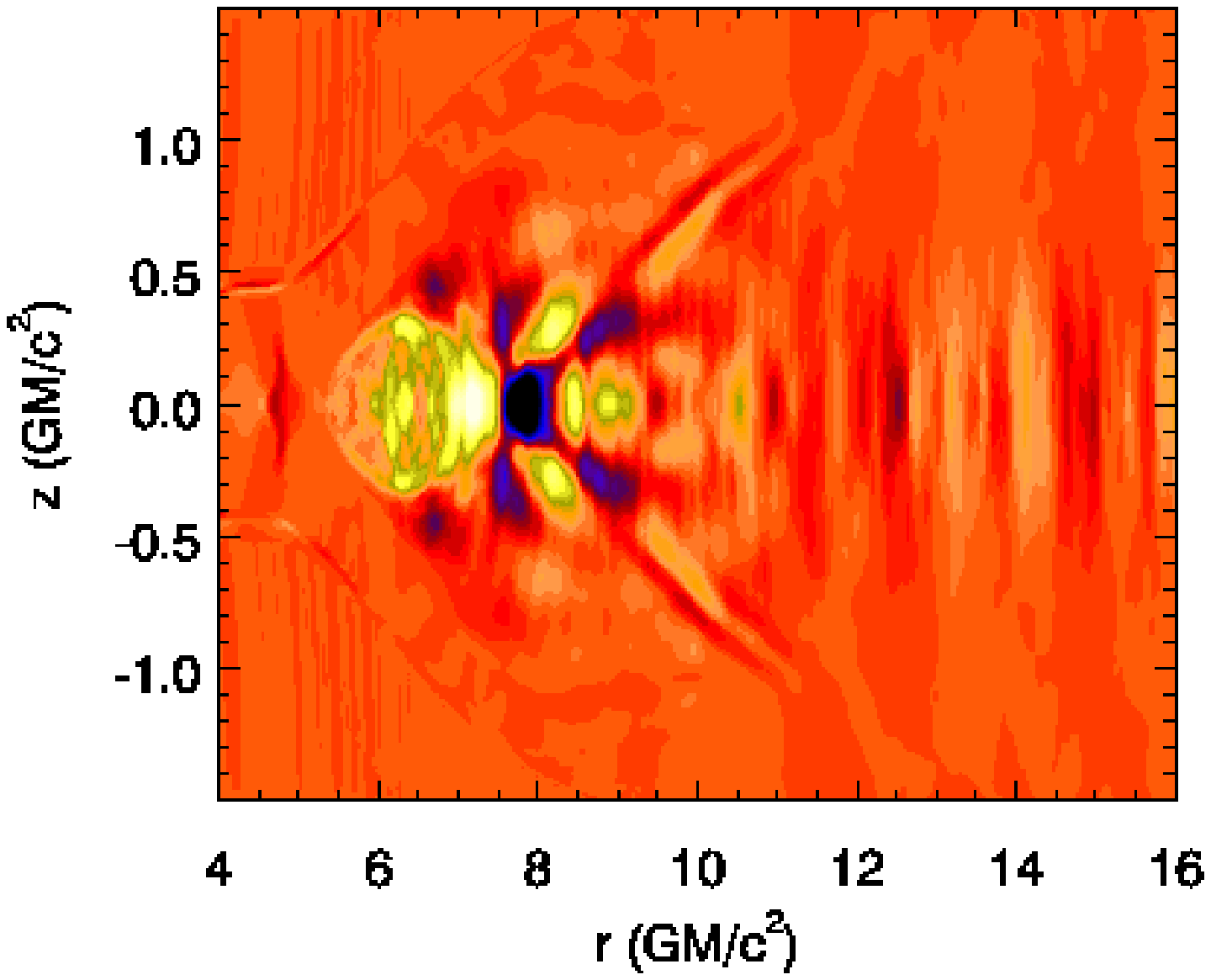,width=0.55\textwidth}
\psfig{figure=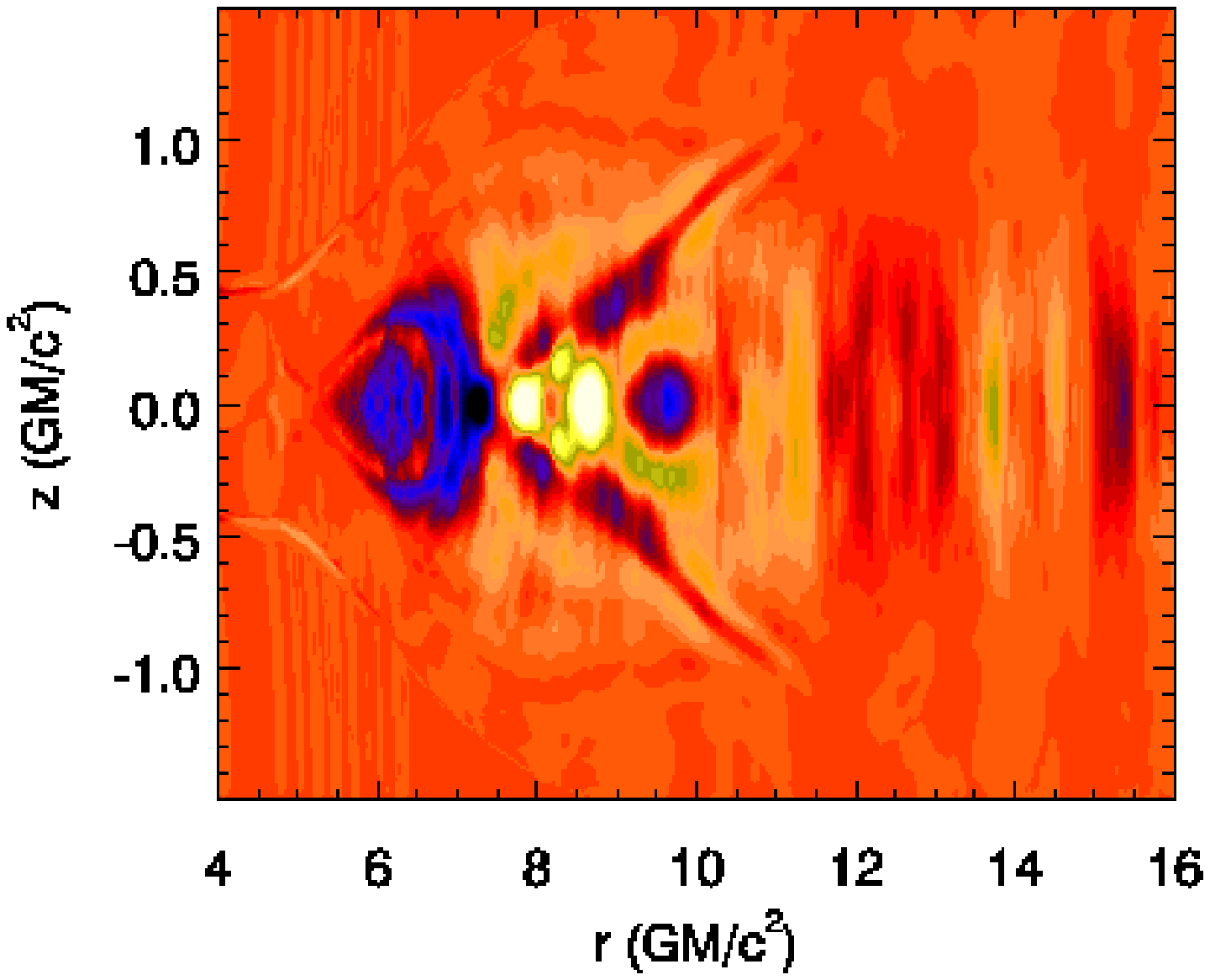,width=0.55\textwidth}
}
\caption{Maps of period-folded pressure deviation (i.e., the
difference between the instantaneous pressure and the time-averaged
pressure) for run HD2d\_1hr.  Only the last 102.4 orbits of data have
been included in order to avoid the transient behavior associated with
the initial conditions.  The folding period corresponds to the peak of
the PSD see in Fig.~\ref{fig:gmode}.  Phases of 0.0 (top-left), 0.25
(top-right), 0.5 (bottom-left) and 0.75 (bottom-right) are shown.
This, in essence, gives us a direct view of the eigenmode.}
\label{fig:folding}
\end{figure*}

Inside of $r=12r_g$, however, the PSDs of these two simulations are
very similar and we can trust that neither the resolution nor the
outer radial boundary affect the results significantly.  We note that
simulations in which the initial radial density profile of the disk is
truncated before reaching the outer boundary also produce essentially
identical results, again giving us confidence that noise infiltration
from the outer boundary is not an important issue.  In addition to
very low frequency noise, the most prominent feature of the inner-disk
PSD is a vertical ridge of enhanced power at $\nu\approx (4-5)\times
10^{-3}\,c^3/GM$ extending from $r\approx 6.5r_g$ out to $r\approx
9.5r_g$.  The one dimensional cuts through this ridge in
Fig.~\ref{fig:gmode} demonstrate it to have all of the expected
properties of a trapped $g$-mode.  Firstly, it exists in a rather
narrow range of frequencies, $\nu\approx (4-5)\times 10^{-3}\,c^3/GM$,
just below the maximum radial epicyclic frequency ($\kappa_{\rm
max}\approx 5.52\times 10^{-3}\,c^3/GM$).  Secondly, it is spatially
bounded by the radii at which the mode frequency becomes equal to the
radial epicyclic frequency.  There is, however, some leaking of the
mode down to the ISCO.

We can visualize the eigenmode by producing maps of pressure
deviations that have been ``period-folded'' on the period
corresponding to the peak power in this mode.  More precisely, we use
the last half of the simulation to produce maps of the difference
between the instantaneous pressure and the time-averaged pressure with
a sampling rate of $0.2T_{\rm isco}$.  We then sort these maps into 16
phase bins (based on the period corresponding to the peak power in
this mode) and average together all maps within a given phase bin.
The result is shown in Fig.~\ref{fig:folding}.  In addition to the
g-mode itself, these maps reveal a striking ``chevron'' pattern at
large radii.  Time-sequences of maps reveal these features to be
outward-radially traveling acoustic waves driven by the global
g-mode, refracted into the upper layers of the disk atmosphere as they
propagate.

\begin{figure}[t]
\centerline{
\psfig{figure=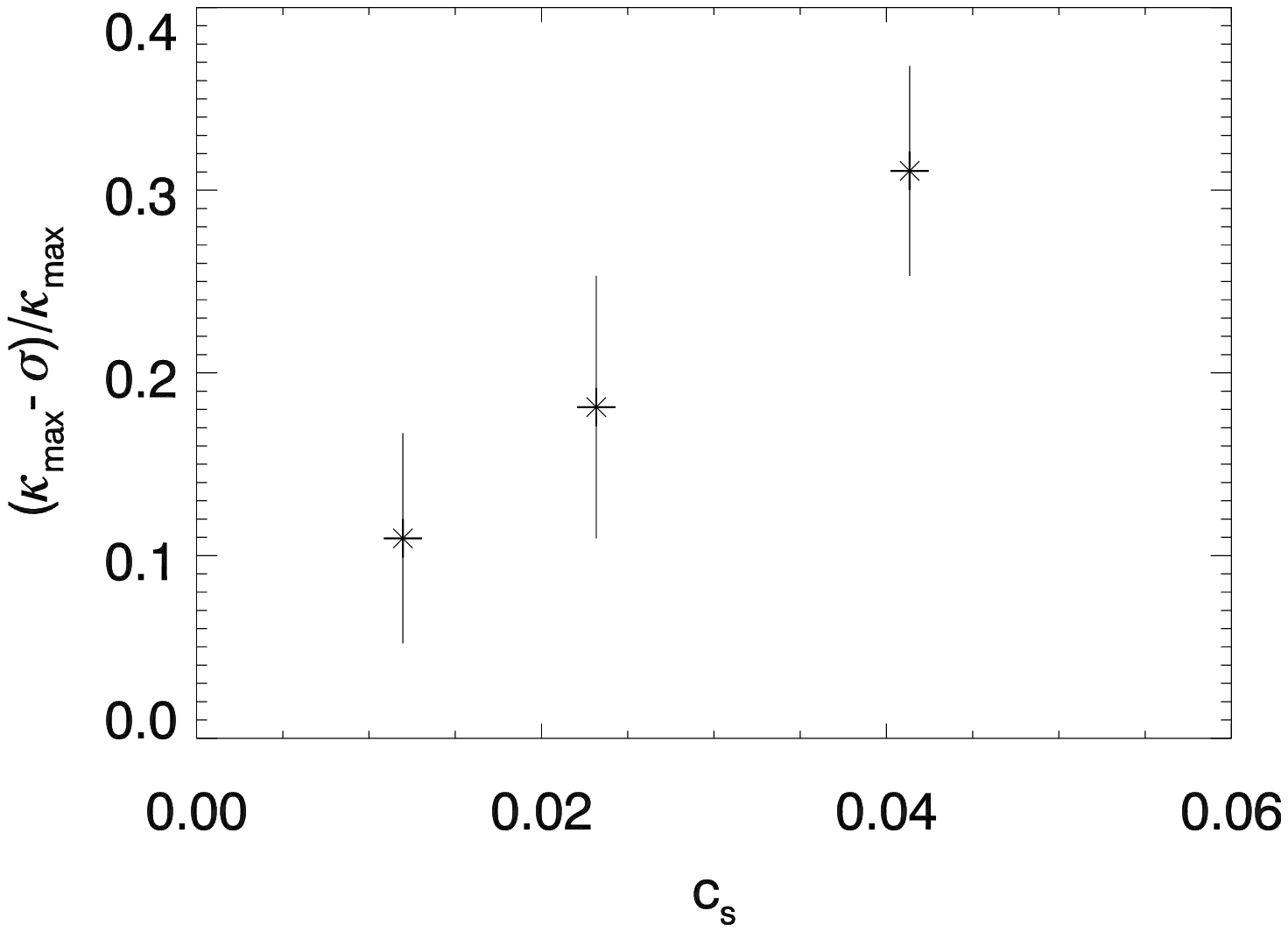,width=0.55\textwidth}
}
\caption{Frequency of the g-modes as a function of mid-plane sound
speed at $r=8r_g$ for runs HD2d\_2, HD2d\_1, and HD2d\_3 (from left to
right).  The vertical bars indicate the range of frequencies over
which enhanced power is seen.}
\label{fig:f_vs_cs}
\vspace{0.5cm}
\end{figure}

As a final demonstration that we have properly identified trapped
$g$-modes in our axisymmetric hydrodynamic simulations, we use our set
of runs to study the dependence of the mode frequency on the sound
speed in the disk.  Figure~\ref{fig:f_vs_cs} shows the dependence of
the mode frequency on midplane sound speed of the disk (and hence disk
thickness) at $r=8r_g$.  As expected from analytic theory (see
Appendix~A), the difference between the mode frequency and the maximum
epicyclic frequency depends linearly on the sound speed.  There is,
however, a discrepancy in the slope of this linear relationship obtained
by the simulations and expected from the analytic theory.  Note that
the analytic treatment assumes that the gas is strictly isothermal
across the whole region of interest, a condition that is clearly
violated in the simulated disk.  Given the sensitivity of the mode to
the radial structure of the disk (as demonstrated by the factor of 2.5
difference in the analytic results between the two pseudo-Newtonian
potentials examined in Appendix~A), the non-isothermality of the gas
in the simulated disk can readily shift the mode frequency away from
the analytic value.

\subsection{Extension to 3-dimensional hydrodynamic models}
\label{sec:3dhydro}

We begin our analysis of the 3-dimensional hydrodynamic simulations
(HD3d\_1) by examining the decay of hydrodynamic perturbations in our
canonical 3-d run (HD3d\_1) using the quantity $K(t)$ as defined in
eqn.~\ref{eq:decay}.  The $K(t)$ behavior for HD3d\_1 is somewhat
different from the axisymmetric simulations at late times, reaching a
quasi-steady state at $K/\max(K)\sim 3\times 10^{-3}$ with aperiodic
fluctuations rather than continuing a ringing exponential decay
(Fig.~\ref{fig:hydro_decay}).  Numerical dissipation must be just as
effective in HD3d\_1 as compared with HD2d\_1, hence these
perturbations must be driven by some instability.  While it is thought
that free Keplerian accretion disks are stable to linear hydrodynamic
perturbations (see Balbus \& Hawley 1998; Hantao et al. 2006), the
presence of the simulation boundaries can introduce true instabilities
(associated with reflection from the boundaries) as well as
uncontrolled numerical noise, and these might explain these low level
sustained fluctuations.  Visual inspection of density slices in the
$(r,z)$ and $(r,\phi)$ planes also suggests that non-axisymmetric
wave-like perturbations interacting with the boundaries are
responsible for perturbing the 3-d hydrodynamic disk.  However, a
detailed study of the sustained fluctuations in the 3-d hydrodynamic
disks is beyond the scope of this paper.  Since these perturbations
exist at a low level (particularly compared with the MHD turbulence
discussed in \S~4; see Fig.~\ref{fig:hydro_decay} for a comparison of
the sustained fluctuations in the 3-d hydrodynamic run compared with
the canonical MHD run) and are likely to be driven by our boundaries
(hence, are not of astrophysical relevance), they are not of
importance for the principal focus of our study.

\begin{figure}[b]
\centerline{
\psfig{figure=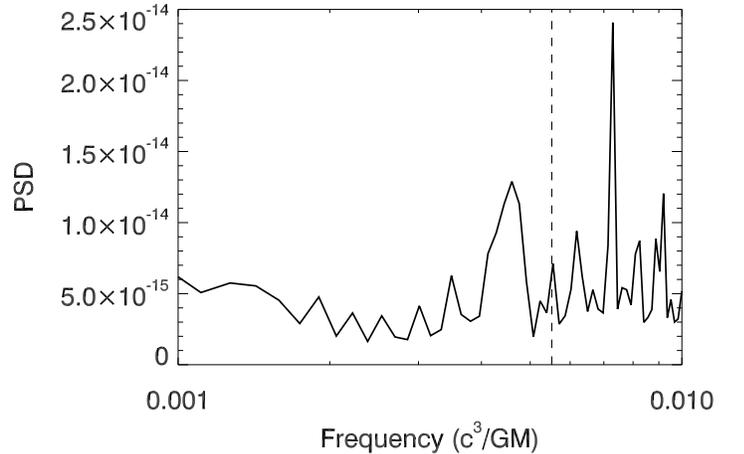,width=0.55\textwidth}
}
\caption{
Midplane PSD of azimuthally-averaged pressure for run HD3d\_1 summed
up over a range of radii $\Delta r=0.5r_g$ centered on $r=8r_g$.  The
dashed line shows the maximum epicyclic frequency.  
}
\label{fig:hd3d_psd}
\end{figure}

The principal issue to be addressed here is the impact of the
transition to 3-dimensions on the presence of the trapped $m=0$
g-modes in the simulated disks.  We might expect the power in these
axisymmetric modes to be reduced in the 3-d case as the free energy is
shared with non-axisymmetric modes.  This is indeed the
case. Figure~\ref{fig:hd3d_psd} shows the PSD at $r=8r_g$ of the last
$\Delta t=102.4T_{\rm isco}$ of the canonical 3-d run HD3d\_1.  A
region of enhanced power is clearly seen in the correct range of
frequencies $\nu\approx (4-5)\times 10^{-3}\,c^3/GM$ to be identified
with the axisymmetric g-modes studied in \S~\ref{sec:axisym}.
Furthermore, examination of the radially-resolved PSD shows that this
region of enhanced power is bounded by the radial epicyclic frequency
in precisely the manner expected for trapped g-modes.  A comparison of
the absolute values of the PSD across the mode does reveal, however,
that the mode contains almost an order of magnitude less power than in
the axisymmetric case.   

We note the existence of a narrow but large amplitude spike just above
a frequency of $7\times 10^{-3}\,c^3/GM$ in the $r=8r_g$ PSD of this
run.  The identification of this feature is not clear; it is not at
the frequency of any expected global g- or p-mode.  It is, however,
confined to a single frequency bin, contains only a small amount of
power and only shows up over a narrow range of radii.  It seems likely
that this is a noise spike.

\section{Magnetohydrodynamic disks}
\label{sec:mhd}

Having gained an understanding of the thin hydrodynamic disks, we move
onto the more astrophysically relevant case of MHD disks.  In this
section, we construct MHD simulations of thin accretion disks in a PW
potential.  We then examine the properties of the broad-band noise and
search for modes in the resulting turbulent MHD disks.

\subsection{Simulation set-up}

We simulate geometrically-thin MHD accretion disks by building upon
our hydrodynamic computational set-up described in
\S~\ref{sec:hydro_setup}.  As discussed in \S~\ref{sec:initial_hydro},
we only consider 3-dimensional MHD simulations.

Table~1 details our set of MHD simulations.  Most of our discussion
will center around run MHD\_1 (which we shall refer to as our
canonical MHD run).  In this run, the hydrodynamic variables are
set-up as for the canonical hydrodynamic run (see
\S~\ref{sec:hydro_setup}) except that we begin the disk in vertical
hydrostatic equilibrium ($h_1=h_2=0.3r_g$ corresponding to
$h_1/r=h_2/r=0.05$ at the ISCO).  An initially weak magnetic field is
introduced in the form of poloidal field loops specified in terms of
their vector potential ${\bf A}=(A_r,A_z,A_\phi)$ in order to ensure
that the initial field is divergence free.  We choose the explicit
form for the vector potential,
\begin{eqnarray}
A_\phi=A_0\,f(r,z)\,p^{1/2}\,\sin \left(\frac{2\pi r}{5h_1}\right),\hspace{0.5cm}A_r
=A_z=0,
\end{eqnarray}
where $A_0$ is a normalization constant and $f(r,z)$ is an envelope
function that is unity in the body of the disk and then smoothly goes
to zero at $r=r_{\rm isco}$, $r=r_{\rm out}$ and at a location three
pressure scale heights away from the midplane of the disk.  The use of
$f(r,z)$ keeps the initial field configuration well away from either
the radial boundaries of the initial disk configuration or the
vertical boundaries of the calculation domain.  The final
multiplicative term produces field reversals with a radial wavelength
of $5h$.  This results in a number of distinct poloidal loops
throughout the disk.  The normalization constant $A_0$ is set to give
an initial ratio of gas-to-magnetic pressure of $\beta=10^3$ in the
inner disk.  In our canonical MHD simulation, this initial condition
is evolved for a duration of $630T_{\rm isco}$ ($38800\,GM/c^3$) using
the MPI-parallelized version of ZEUS described above.

We supplement the ideal MHD algorithms of ZEUS in two ways.  Firstly,
it is necessary to impose a floor to the density field of $10^{-5}$
times the initial maximum density in order to prevent the numerical
integration from producing negative densities.  This essentially
amounts to a subtle distributed mass source.  The density only reaches
this floor close to the $z$-boundary (i.e., many scale heights above
and below the disk plane).  Secondly, we implement the prescription of
Miller \& Stone (2000) to include some effects of the displacement
current, principally forcing the Alfv\'en speed to correctly limit to
the speed of light as the magnetic fields becomes strong.  We note
that this ``Alfv\'en speed limiter'' only plays a role within small
patches of the tenuous magnetized atmosphere that forms at large
vertical distances above and below the disk; it never plays a
significant role in the body of the accretion disk.

Periodic boundary conditions were imposed on the $\phi$-boundaries,
and zero-gradient outflow boundary conditions were imposed at both
the inner and outer radial boundaries (Stone \& Norman 1992a,b).
However, the choice of the $z$-boundary condition for this kind of
simulation is notoriously problematic.  The most physically motivated
choice would be a free outflow boundary.  However, as described in
Stone et al. (1996), field-line ``snapping'' at these free boundaries
can halt such a simulation.  Indeed, our own test simulations
employing zero-gradient outflow boundary conditions on the
$z$-boundaries were found to be subject to these difficulties, as well
as occasional numerical instabilities appearing to result from an
interplay of the imposition of the density floor and the free
boundary.  Furthermore, these tests showed that the tenuous matter
high above the disk mid-plane generally flows slowly across these
boundaries at very sub-sonic and sub-Alfv\'enic speeds; strictly, this
invalidates the use of such boundary conditions anyways (since the
flow on the other side of the boundary should be able to act back on
the simulation domain).

We adopt the solution of Stone et al. (1996) and choose to employ
periodic boundary conditions in the $z$-directions.  While this is
obviously unphysical in the sense that matter cannot leave the
simulation domain in the vertical direction, it does guarantee
mathematically reasonable behavior at the boundary (eliminating
numerical instabilities) and, more importantly, appears to have no
effect on the dynamics of the accretion disk itself (as diagnosed
through comparisons with our vertical-outflow test runs).  In order to
further isolate the simulated disk from the vertical boundaries, we
expand the vertical domain (compared to the canonical hydrodynamic
simulation) to $z\in(-3,3)$ (i.e., $\pm 10h_1)$.

We perform four additional simulations aimed at demonstrating the
robustness of the canonical simulation.  In run MHD\_2, we restrict
the vertical domain back to $z\in(-1.5,1.5)$, allowing us to gauge the
importance of the location of the $z$-boundaries.  Run MHD\_2hr has an
identical set-up to MHD\_2 except that the radial and vertical
resolution is doubled compared with the canonical run (i.e., the voxel
size is reduced by a factor of two in each of the radial and vertical
directions), and the azimuthal domain is doubled to
$\Delta\phi=60^\circ$ at fixed resolution.  Due to the factor of 8
increase in number of computational cells (and the decrease in the
timestep), this run was only integrated for a duration of $85T_{\rm
isco}$ ($5236\,GM/c^3$).  While the run time is insufficient to
conduct a detailed temporal study, a comparison with run MHD\_2 does
allow us to investigate the effect of both the radial/vertical
resolution and the extent of the $\phi$-domain on the turbulent state
(see below).  As discussed below, we find that the properties of the
simulated disks are very similar these two runs.

Run MHD\_3 is similar to MHD\_2 except that the outer radius is pushed
to $r=28r_g$ (at fixed resolution), doubling the size of the radial
domain.  Comparing MHD\_2 and MHD\_3, we find no evidence that the
outer radial boundary is affecting the inner disk ($r<12r_g$) in any
way.

\subsection{Basic evolution of the MHD disks}
\label{sec:basicmhdevol}

\begin{figure*}
\hbox{
\hspace{-1cm}
\psfig{figure=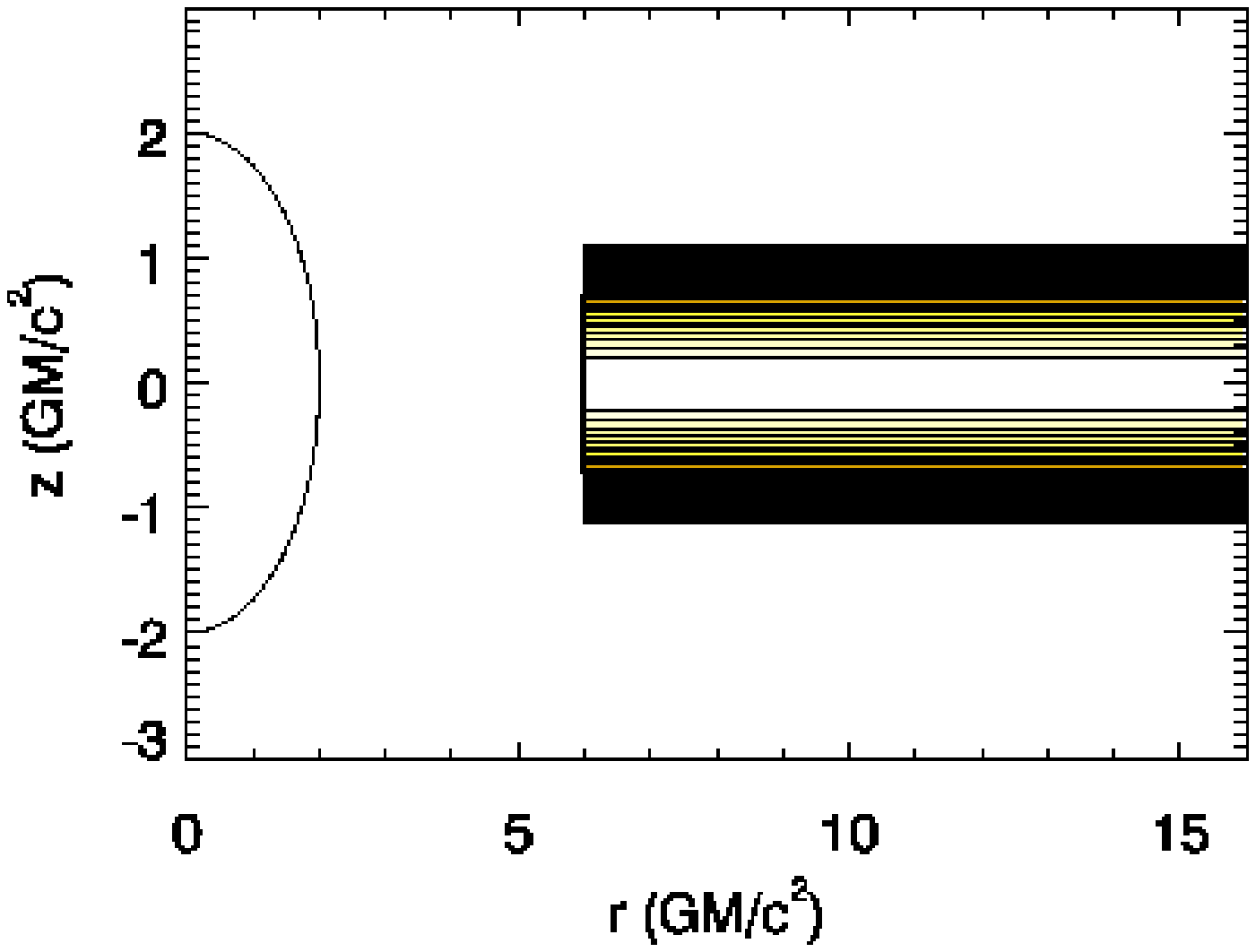,width=0.55\textwidth}
\hspace{-1cm}
\psfig{figure=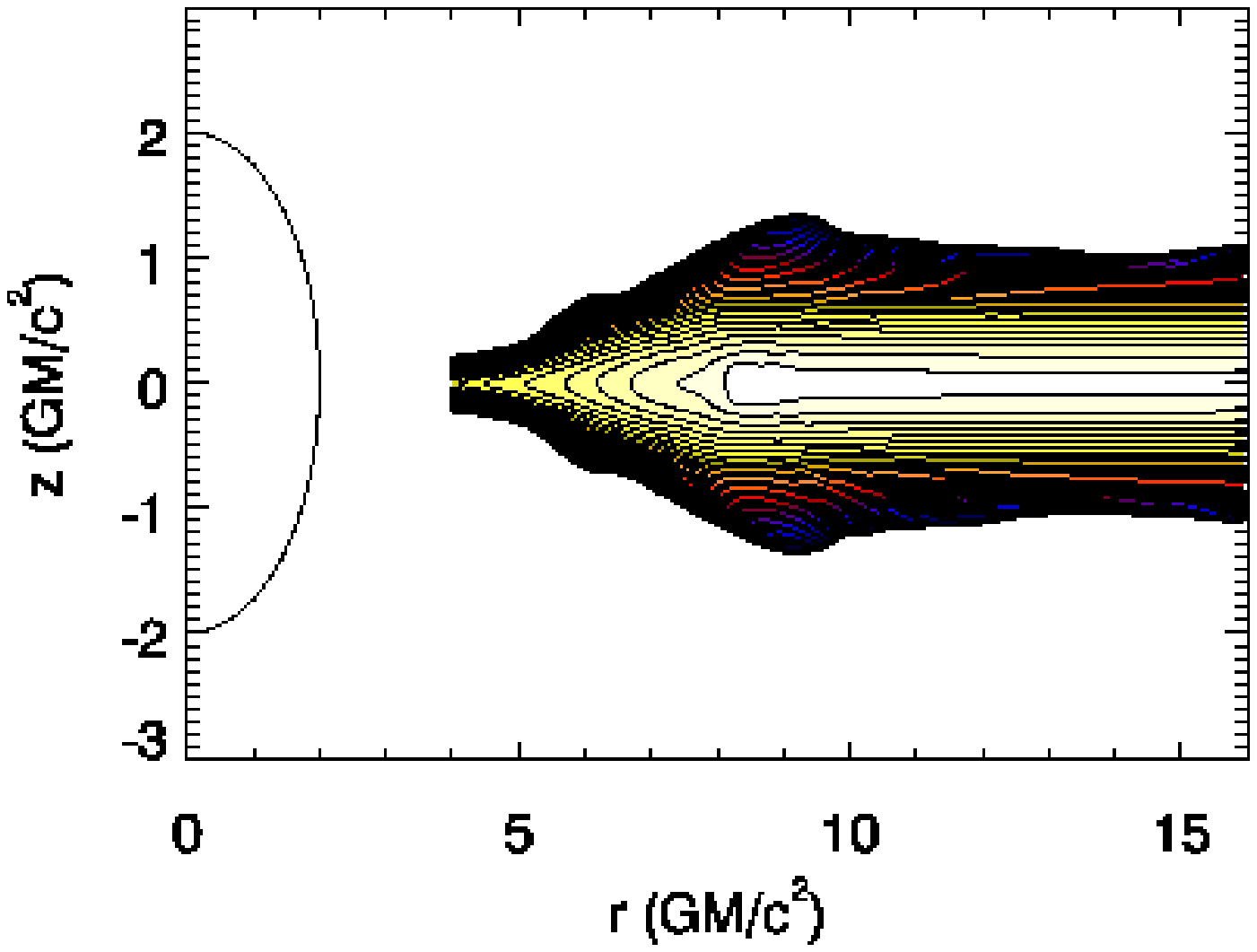,width=0.55\textwidth}
}
\hbox{
\hspace{-1cm}
\psfig{figure=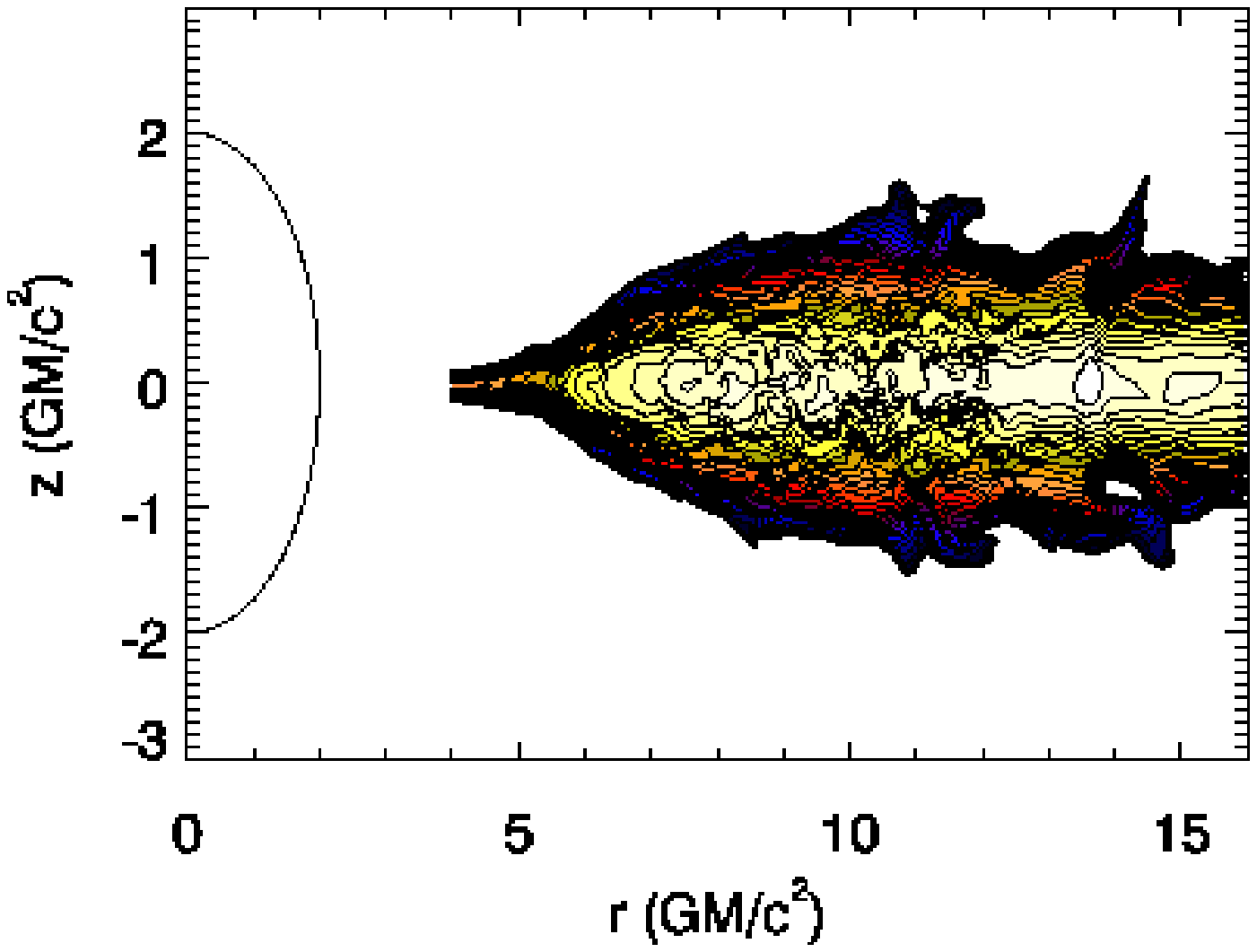,width=0.55\textwidth}
\hspace{-1cm}
\psfig{figure=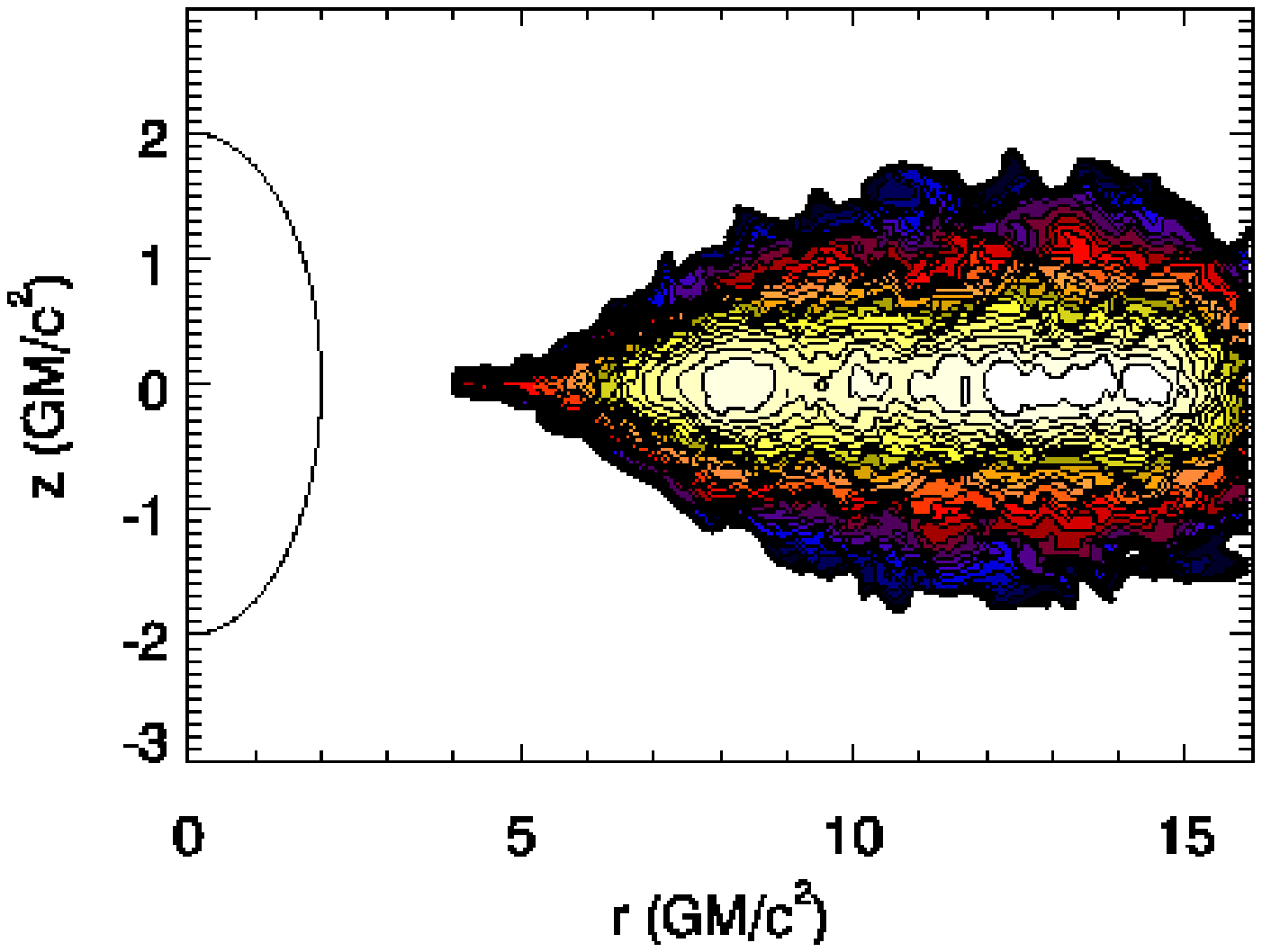,width=0.55\textwidth}
}
\caption{Snapshots of evolution of the canonical MHD simulation
(MHD\_1) at $t=0$ (top-left), $t=1T_{\rm isco}$ (top-right),
$t=10T_{\rm isco}$ (bottom-left) and $t=100T_{\rm isco}$
(bottom-right), where $T_{\rm isco}$ is the orbital period at the
ISCO.  Both the color table and contours show the logarithmic density
structure of the disk cross-section, with 10 contours per decade of
density.  A range of densities spanning three orders of magnitude are
shown.  The curved line to the left of each frame represents the event
horizon.}
\label{fig:mhd_evol}
\vspace{0.5cm}
\end{figure*}

We now discuss the evolution and general properties of these MHD
disks, centering our discussion around run MHD\_1.  At very early
times ($t<5T_{\rm isco}$) strong hydrodynamic transients dominate the
evolution as radial pressure forces drive mass into the region within
the ISCO.  Similarly to the hydrodynamic cases, these transients
launch outwardly directed axisymmetric waves that break to form rolls.
These strong hydrodynamic transients largely damp away in all but the
outermost parts of the disk within $10T_{\rm isco}$.  Concurrently,
the MRI amplifies the initial magnetic field until the (domain wide)
volume-averaged ratio of gas-to-magnetic pressure peaks at
$\langle\beta\rangle \sim 5$ (at $t\approx 10T_{\rm isco}$), at which
point most of the flow has become turbulent.  The entire flow (outside
of the ISCO) becomes turbulent by $t\approx 20T_{\rm isco}$.
Figure~\ref{fig:mhd_evol} displays the density field across a vertical
slice through the accretion disk at various times.

\begin{figure}
\centerline{
\hspace{-1cm}
\psfig{figure=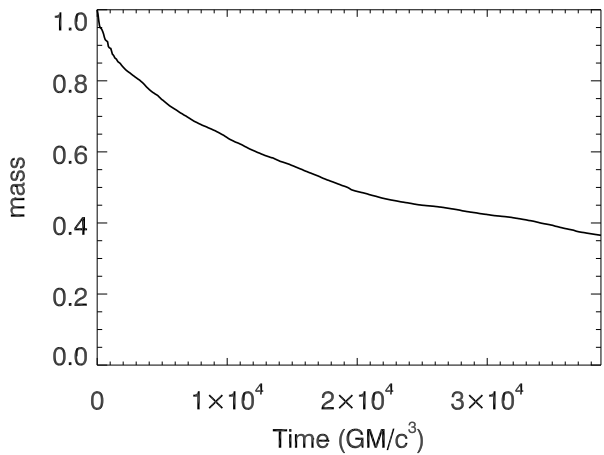,width=0.55\textwidth}
}
\caption{Normalized total mass as a function of time for run MHD\_1.}
\label{fig:mass_loss}
\vspace{0.5cm}
\end{figure}

Between $t=10T_{\rm isco}$ and $t=20T_{\rm isco}$, the total magnetic
energy in the computational domain declines until $\langle\beta\rangle
\sim 20$.  After this, a quasi-steady state seems to be achieved where
the magnetic field generation by the MRI-driven MHD turbulence is
balanced by the removal of magnetic field energy due to numerical
reconnection and magnetic buoyancy.  The fact that buoyancy is playing
an important role is revealed by examining time-sequences of the
strengths of B-field components in $(r,z)$ cross-sections of the disk.
One clearly sees highly magnetized structures being generated close to
the midplane of the disk which then propagate vertically away from the
midplane\footnote{In principle, one could address the importance of
magnetic buoyancy in removing magnetic energy from the mid-plane
regions of the disk by comparing the vertical Poynting flux with
(numerical) reconnection losses.  However, energy losses due to
numerical reconnection are cannot be tracked in our simulation (due to
the non-conservative nature of ZEUS algorithm) and, hence, a rigorous
study of this issue is not possible.}.  From this time until the end
of the simulation at $t=630T_{\rm isco}$, MHD turbulence and hence
accretion is sustained.  Over the course of the simulation, the disk
loses a little more than 60\% of its mass (Fig.~\ref{fig:mass_loss}).

\begin{figure*}
\hbox{
\hspace{-1cm}
\psfig{figure=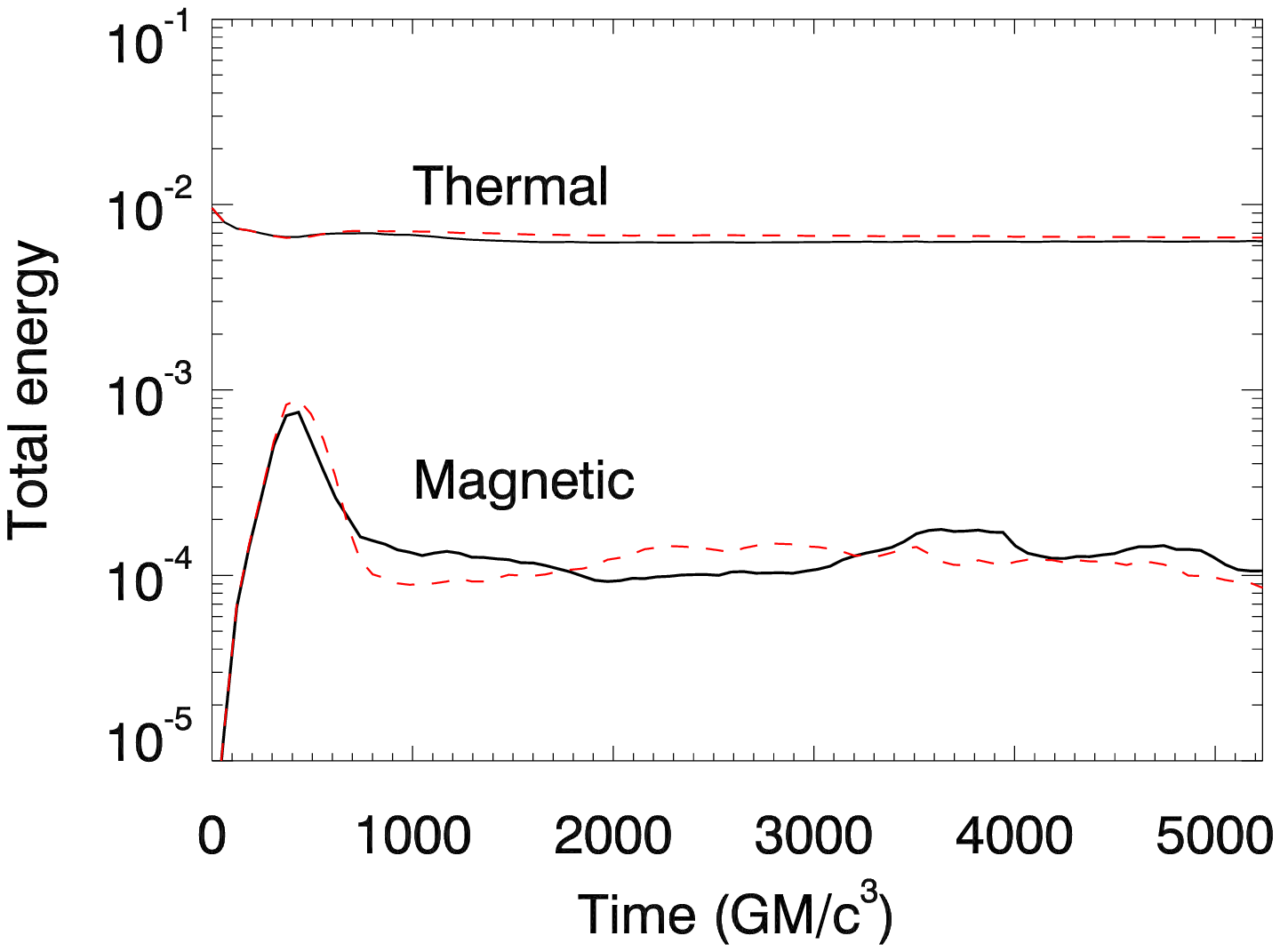,width=0.55\textwidth}
\hspace{-1cm}
\psfig{figure=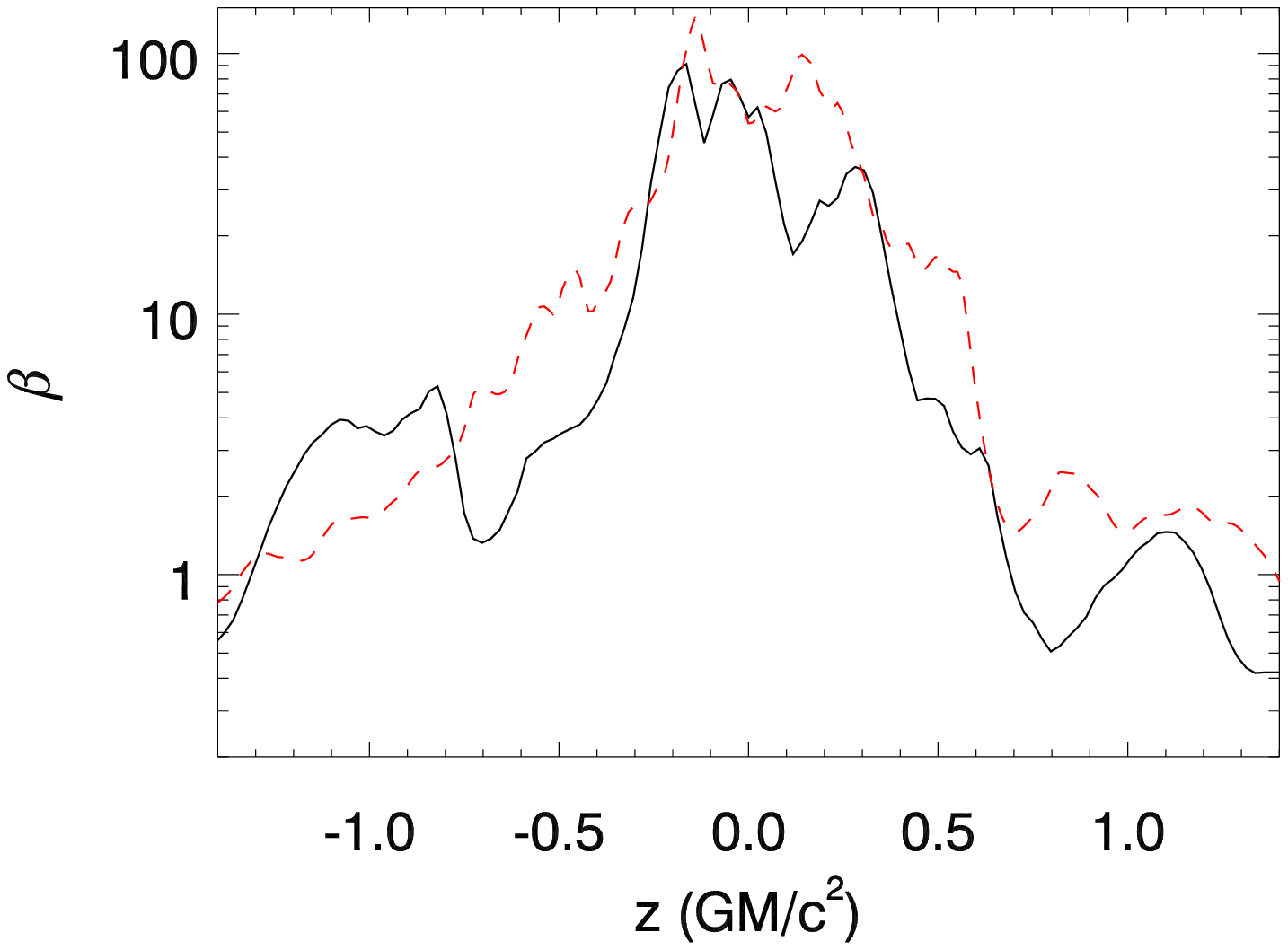,width=0.55\textwidth}
}
\caption{{\it Left panel : }Total (domain integrated) magnetic and
thermal energies for run MHD\_2 (black solid line) and its
high-resolution counterpart (MHD\_2hr; red dashed line).  {\it Right
panel : }Azimuthally-averaged plasma-$\beta$ parameter (i.e., the
ratio of the thermal to magnetic pressure) as a function of vertical
height in the disk at $r=8\,r_g$. To obtain this plot, data from
$t=30-35T_{\rm isco}$ have been averaged together.  The solid (black)
line shows run MHD\_2 whereas the dashed (red) line shows its
high-resolution counterpart (MHD\_2hr).}
\label{fig:mhd_compare}
\end{figure*}

The total magnetic energy and thermal energy undergo a slow decline as
mass is drained out of the simulated disk.  During this decline, the
volume-averaged plasma-$\beta$ parameter within the domain remains in
the range $\langle\beta\rangle\sim 20-30$.  However, as expected, the
$\beta$ parameter within the high-density body of the disk is
appreciably higher, reaching values of $\beta\approx 70-100$.  While
this is significantly larger than the $\beta$ found in global
simulations of thicker disks (e.g., Hawley \& Krolik 2001, 2002), it
is in line with what might be expected for thin accretion disks with
zero net field as diagnosed through local shearing-box simulations
both with and without vertical stratification (Stone et al. 1996;
Hawley, Gammie, \& Balbus 1996; Miller \& Stone 2000).

A generic concern in this class of simulation is the affect of the
$z$-boundaries.  Once it achieves its quasi-steady state, our
canonical MHD simulation displays a 2--3 order of magnitude drop in
magnetic pressure, and a 5 order of magnitude drop in gas pressure,
between the disk and the $z$-boundary.  Thus, the disk boundary seems
to be well isolated from the boundary.  Further confidence is gained
from an examination of run MHD\_2 in which the $z$-boundaries have
been brought in from $z=\pm3$ to $z=\pm1.5$.  Despite the fact that
the magnetic pressure now only drops by 1 order of magnitude from the
disk to the boundary, all of the results from the canonical MHD run
discussed in this paper are reproduced by MHD\_2.  More precisely, (1)
the PSD of the fluid variables (\S\ref{sec:mhdfluidpsd}) are
essentially indistinguishable, failing to show any evidence for global
modes but displaying prominent local $p$-modes, (2) the PSD of the
mass accretion rate has a broken power-law form with indices that
differ from those found in MHD\_1 by $\Delta\approx 0.1$
(comparable to the 1$\sigma$ error bars) and break frequencies that
differ by $\Delta (\log \nu_{\rm break})\approx 0.05$ (again,
comparable to the 1$\sigma$ error bars).

Another generic concern with thin disk simulations is whether the
resolution in the vertical direction is adequate.  The canonical MHD
simulation (MHD\_1) has approximately 13 grid cells spanning a
vertical range $\Delta z=h_2$, implying that we cannot follow any
modes with a wavelength smaller than $\lambda\sim h_2/2$.  This is
just sufficient to follow the fastest growing MRI mode (with
wavelength $\lambda\sim 2\pi h/\beta^{1/2}$) if $\beta\approxlt 100$.
To ensure that we have, in fact, achieved adequate resolution, we
compare runs MHD\_2 and its high resolution counterpart, run MHD\_2hr.
For example, Fig.~\ref{fig:mhd_compare} compares the time-dependence
of the thermal and magnetic energies, as well as the height dependence
of the plasma-$\beta$ parameter.  While there is some indication of
increased buoyancy driven escape of magnetic fields from the
high-resolution simulation (as is apparent from the higher value of
$\beta$ at intermediate heights), the two simulations generally
compare very well.  Thus, we conclude that we have achieved adequate
numerical resolution.

\subsection{Temporal power spectra of the basic fluid variables}

\subsubsection{The importance of correcting for secular changes 
during the simulation}
\label{sec:decay_corr}

The length of our canonical MHD run makes it well suited for a detailed
study of temporal variability.  In particular, the long stream of
simulation data facilitates the construction of PSDs.  However,  as we
now discuss, significant complications arise in the analysis of these
MHD simulation as compared with the pure hydrodynamic simulations.

In the case of the hydrodynamic simulations, the background (i.e.,
unperturbed) flow achieves almost a stationary state once the large
transients caused by the initial conditions have died out.  In
particular, the lack of angular momentum transport within the
hydrodynamic disk (other than that due to the very small numerical
viscosity) allows the disk to achieve a non-accreting state.  Density
and pressure fluctuations about this background state can then be
studied via the straightforward construction of the PSD.  The constant
background level does not contribute to the power spectrum and hence
the PSD faithfully characterizes the fluctuations of interest.

However, even once the initial transients have been dissipated, MHD
disks never achieve this kind of stationary background flow.  MHD
turbulence leads to continued accretion that depletes mass from the
simulated disks.  This decline in total mass leads to secular changes
(with an approximately exponential form) in the density and pressure
of the background flow.  Unless corrected for, even a rather slow
exponential decay can have a significant influence on the PSD of the
pressure or density fluctuations, severely affecting attempts to
characterize the properties of the astrophysically relevant
fluctuations (i.e., the fluctuations that would be present in the
ideal case of a steady-state disk in which the mass was replaced from
a large reservoir).

To see this, consider some quantity whose time-series $f(t)$ we extract
from the simulation (for example, this could be the mid-plane
density or pressure at some given radius).  Let us assume that this
can be decomposed as
\begin{equation}
f(t)=\epsilon(t)[1+g(t)],
\label{eq:fluc_decomp}
\end{equation}
where $g(t)$ is the time-series of the astrophysically-interesting
fluctuations about some mean state (i.e., $\langle g\rangle=0$), and
$\epsilon(t)$ is a decay function that describes the secular change in
the background state due to the draining of mass from the simulation.
It must be noted that the decomposition given in
eq.~\ref{eq:fluc_decomp} is not completely general; this assumes that
the amplitude of the ``real'' fluctuations is proportional to the mean
background value, and that the properties of the fluctuations
otherwise remain invariant as the background state decays.  This
decomposition would be valid if the decay of the simulated disk simply
amounted to a gradual decline in the density scale of the simulation
while the temperature and velocity structures remained unchanged.  We
shall refer to such (simulated) disks as {\it density-invariant
disks}.  This does indeed appear to describe our simulated MHD disks
(i.e., there is little or no corresponding secular change in disk
thickness or characteristic turbulent velocities).  In this case, the
mass accretion rate will be proportional to the density and the decay
will hence have an exponential form,
$\epsilon=\epsilon_0\,e^{-t/t_0}$.

In the uninteresting case where there are no fluctuations ($g(t)=0,
\forall t$), the observed signal is just $f(t)=\epsilon(t)$, leading
to a Fourier Transform (FT) and PSD given by
\begin{equation}
\tilde{\epsilon}(\omega)=\frac{A_1}{(1/t_0)+i\omega}, \hspace{0.5cm}{\rm and}\hspace{0.5cm}P_\epsilon(\omega)=\frac{A_2}{(1/t_0)^2+\omega^2},
\end{equation}
where $A_1$ and $A_2$ are uninteresting normalization constants.
Thus, for $\omega\gg 1/t_0$, the PSD of the exponential decay goes as
$P_\epsilon(\omega)\propto\omega^{-2}$.

In the more interesting case of non-zero fluctuations, the FT of the
observed signal is,
\begin{equation}
\tilde{f}(\omega)=\tilde{\epsilon}(\omega)+\int
\tilde{\epsilon}(\omega^\prime)\tilde{g}(\omega-\omega^\prime)\,d\omega^\prime,
\label{eq:combined_ft}
\end{equation} 
i.e., the sum of the FT of the exponential decay with the convolution
of the FTs of the decay and the interesting fluctuations.  When the
fluctuations are small compared with the (decaying) background state,
as is the case for the density and pressure, one can see that the PSD
of the observed signal will be dominated by the $1/\omega^2$
associated with the decay.  Even in the case where the fluctuations
are large compared with the decay, the PSD will still be influenced by
the exponential decay due to the convolution term in
eqn.\ref{eq:combined_ft}.  In particular, regions of the ``real'' PSD
which are steeper than $\omega^{-2}$ (including the high-frequency
wing of any QPO or regions above a high-frequency break) will tend to
get filled.

Clearly, we must correct for this decay of the background state, and
be cognizant of manifestations of any remaining uncorrected effects of
this decay.  This procedure plays the same role as the
``pre-whitening'' employed by Schnittman, Krolik, \& Hawley (2006;
hereafter SKH).  We proceed by dividing the observed time series by a
``best-fitting'' exponential decay function.  Here, the time-constant
of the exponential decay $t_0$ is estimated via two methods.  Firstly,
we can set $t_0$ by the requirement that the starting and final values
of the observed series are equal [$f(t=0)=f(t=T)$, where $t=T$
corresponds to the end of the time-series].  This is the procedure
adopted by Schnittman et al. (2006) except that they employ a linear
form for the decay function.  Secondly, we can set $t_0$ via a
least-square fit of an exponential form to the observed time-series.
These methods give very similar values of $t_0$ and similar corrected
PSDs when applied to density or pressure fluctuations, as we shall now
see.

\subsubsection{Power spectra of fluid variables in disk mid-plane}
\label{sec:mhdfluidpsd}

\begin{figure*}
\begin{center}
\hbox{
\psfig{figure=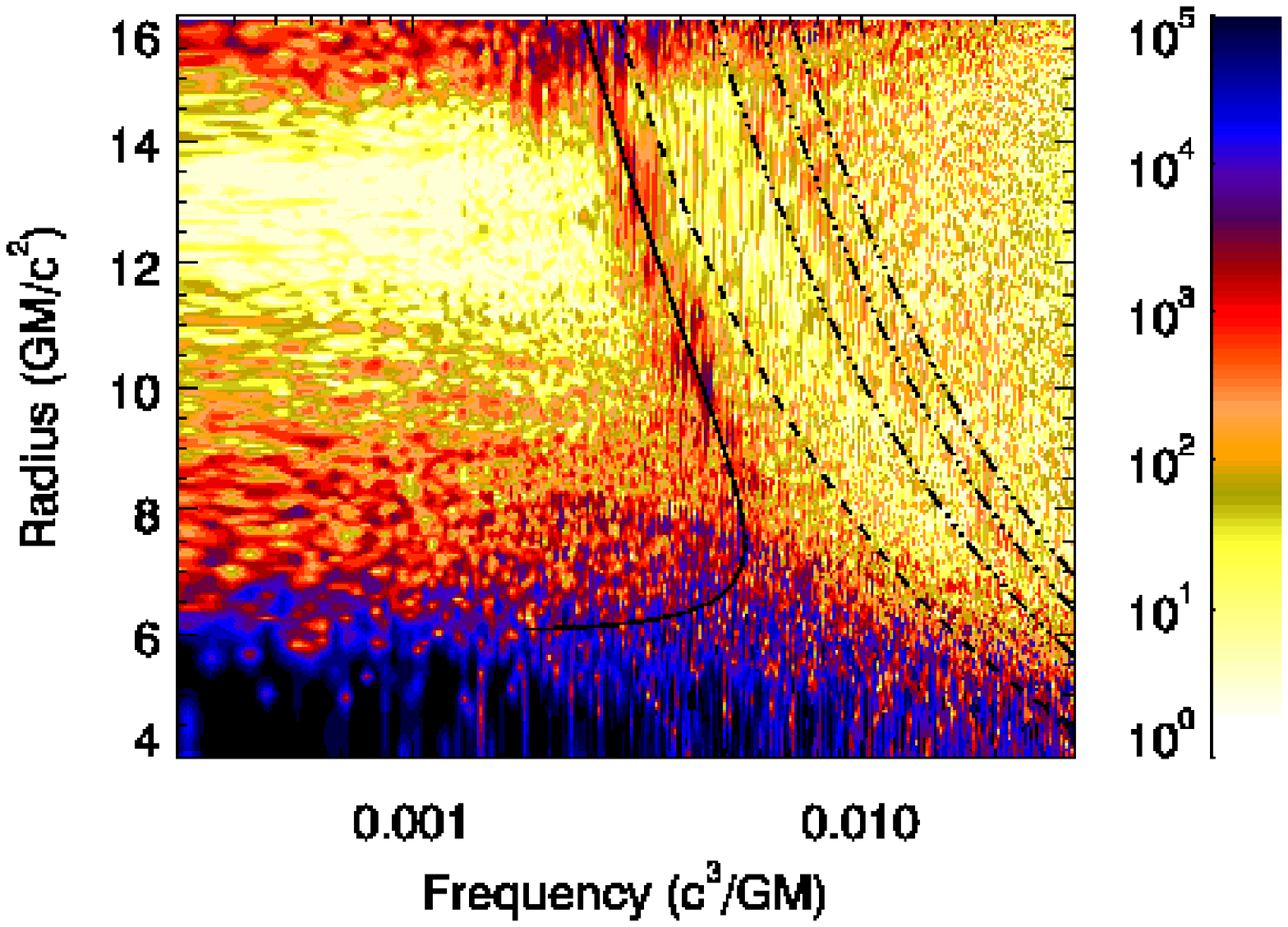,width=0.55\textwidth}
\psfig{figure=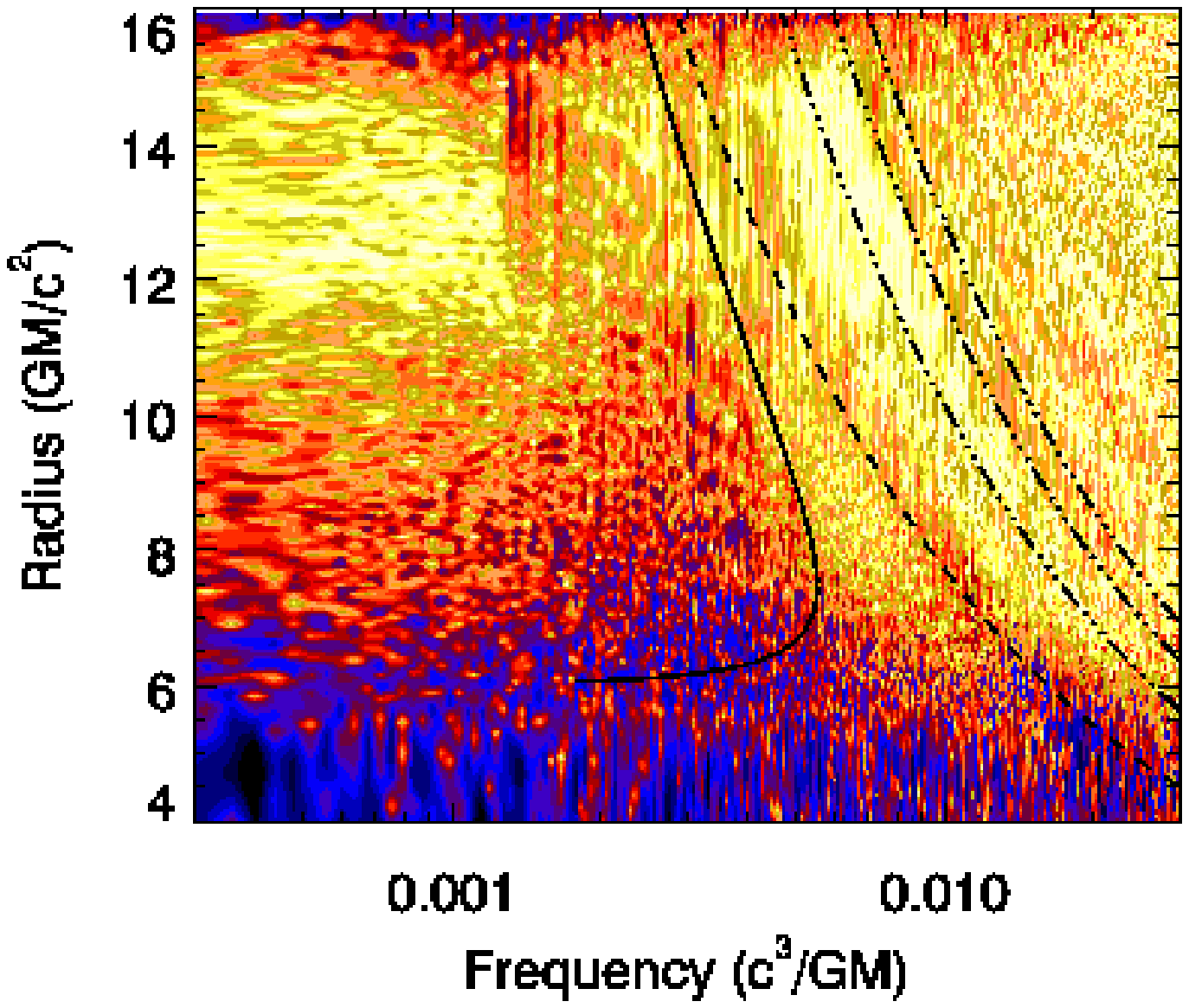,width=0.55\textwidth}
}
\hbox{
\psfig{figure=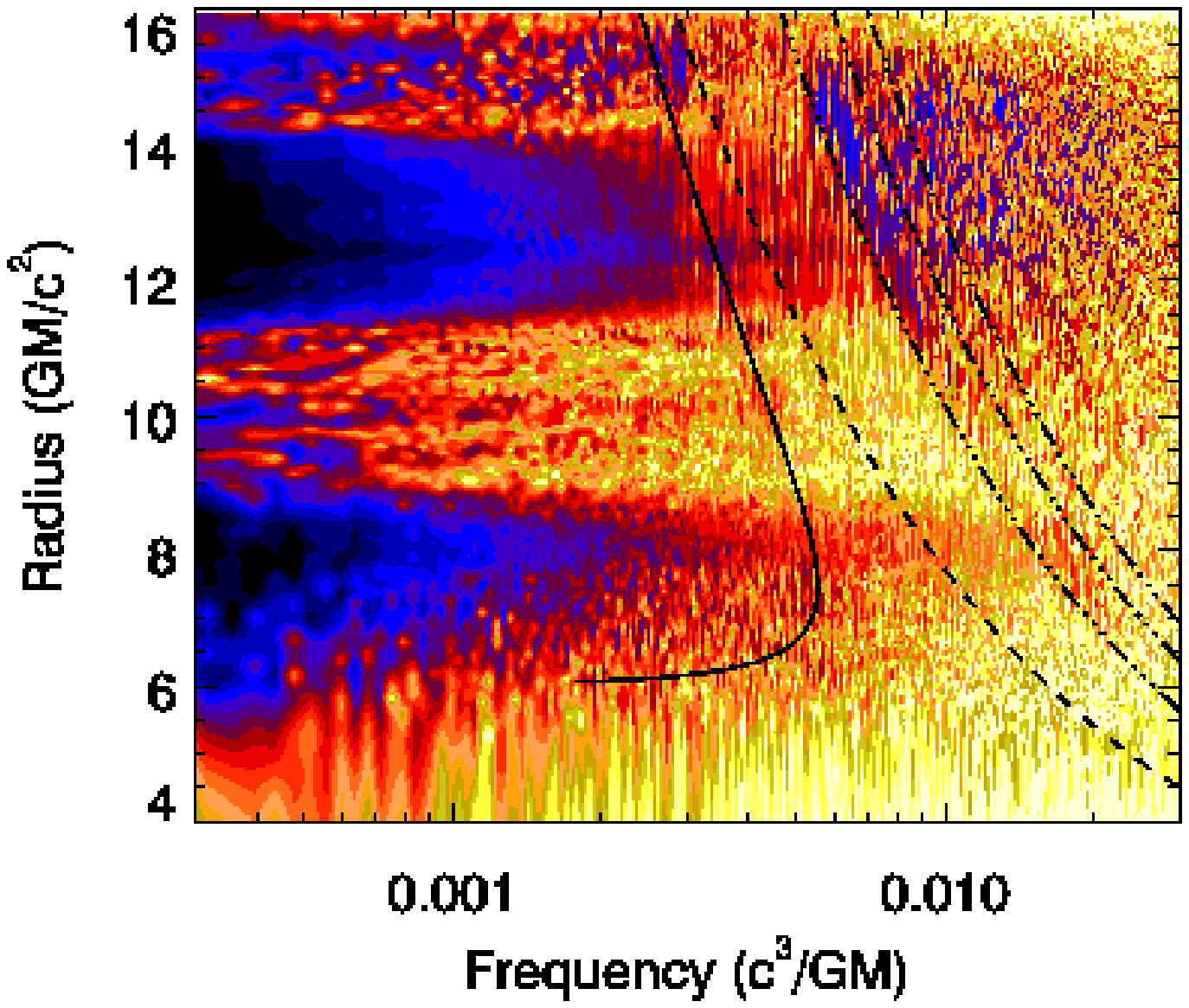,width=0.55\textwidth}
\psfig{figure=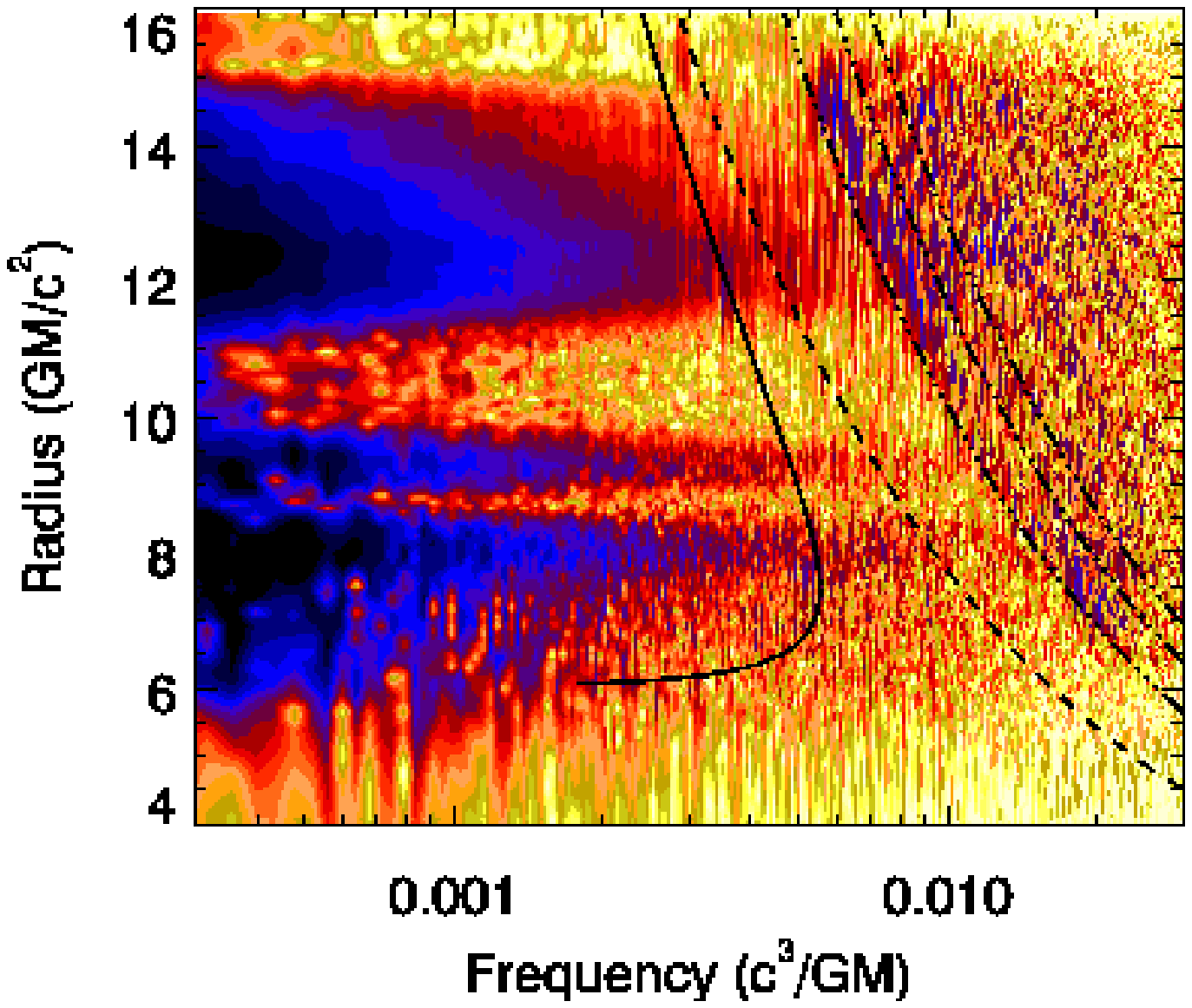,width=0.55\textwidth}
}
\end{center}
\caption{PSD for azimuthally-averaged midplane fluid quantities from
in run MHD\_1.  Panels show PSDs for radial velocity (top left),
vertical velocity (top right), density (bottom left) and pressure
(bottom right).  The density and pressure variables have been divided
by the least-square best-fitting exponential function to correct for
the secular decay of the simulation, as discussed in
\S~\ref{sec:decay_corr}.  Also shown are the radial epicyclic
frequency (solid), orbital frequency (dashed) and the $n=1,2,3$ pure
vertical p-modes (from left to right dot-dashed lines).  The absolute
scaling of the PSDs, as indicated by the color-bar, is arbitrary.  }
\label{fig:mhd_psd}
\end{figure*}

We now examine the PSD of the azimuthally averaged fluid variables
(velocities, pressure and density) for the simulated MHD disks, as we
did with the hydrodynamic disks in \S~\ref{sec:axisym}.  The top
panels of Fig.~\ref{fig:mhd_psd} show the PSD for the radial and
vertical components of velocity in the midplane of the disk, $v^{\rm
mid}_r(r,t)$ and $v^{\rm mid}_z(r,t)$ for a duration lasting $\Delta
t=409.6T_{\rm isco}$ starting at $t=100T_{\rm isco}$ (i.e., well after
all of the initial transients have dissipated and a quasi-steady
turbulent state has been established).  Note that, outside of the
ISCO, these velocity components are themselves first-order fluctuating
quantities (i.e., the radial and vertical velocity of the background
state is approximately zero) and, within the density-invariant disk
assumption, will have characteristic fluctuation amplitudes that
remain constant throughout the simulation.  Hence, we do not need to
correct these PSDs for the effect of the density decay in the
simulation.

It is readily seen that the PSD of the midplane radial velocity
$v_r^{\rm mid}$ is dominated by the radial epicyclic frequency from
$r\sim 7-8r_g$ out to the outer radial boundary.  Inside of
$r\approx 7-8r_g$, the PSD shows broad band power associated with
the transition from turbulent to plunging flow.  

The PSD of the midplane vertical velocity $v^{\rm mid}_z$ is quite
different and can be interpreted in the light of the {\it local} fluid
oscillations discussed in \S~\ref{sec:local_osc}.  Below the radial
epicyclic frequency, there is a broad spectrum of g-modes.  Above the
orbital frequency, the distinct modes described by
eqn.~\ref{eq:vert_modes} become apparent; the $n=0$ mode
($\omega>\Omega$) and $n=2$ mode ($\omega>2.08\Omega$) can be picked
out as distinct tracks, and there are hints of higher frequency modes
as well.  The fact that only even-$n$ modes are seen is readily
understood given that the odd-$n$ modes have $v_z$-nodes at the
midplane, as is evident from the power deficit centered on the $n=1$
frequency in the $v_z$ PSD.

The bottom panels of Fig.~\ref{fig:mhd_psd} show the corresponding PSD
for the density and pressure from the canonical simulation.  The
displayed PSDs have been corrected for the decay of the background
state by dividing through by an exponential curve that best-fits (in a
least square sense) the domain-averaged density or pressure.  Very
similar results are obtained using an exponential determined from just
the end points.  The radial banding of the low-frequency noise is
largely the effect of the pre-whitening procedure; the full
(non-corrected) low-frequency power is somewhat greater and the
bands correspond to the residuals between the real decay and the
simple exponential model.  In addition to this low frequency noise
(which is of secondary interest to the investigation presented in this
paper), both PSDs show an enhancement corresponding to the $n=1$
vertical p-mode (with $\omega>1.63\Omega$).  As expected, there is no
enhancement in the density or pressure PSDs corresponding to the $n=0$
or $n=2$ p-modes; these have pressure and density nodes at the
midplane.

\begin{figure}
\centerline{
\psfig{figure=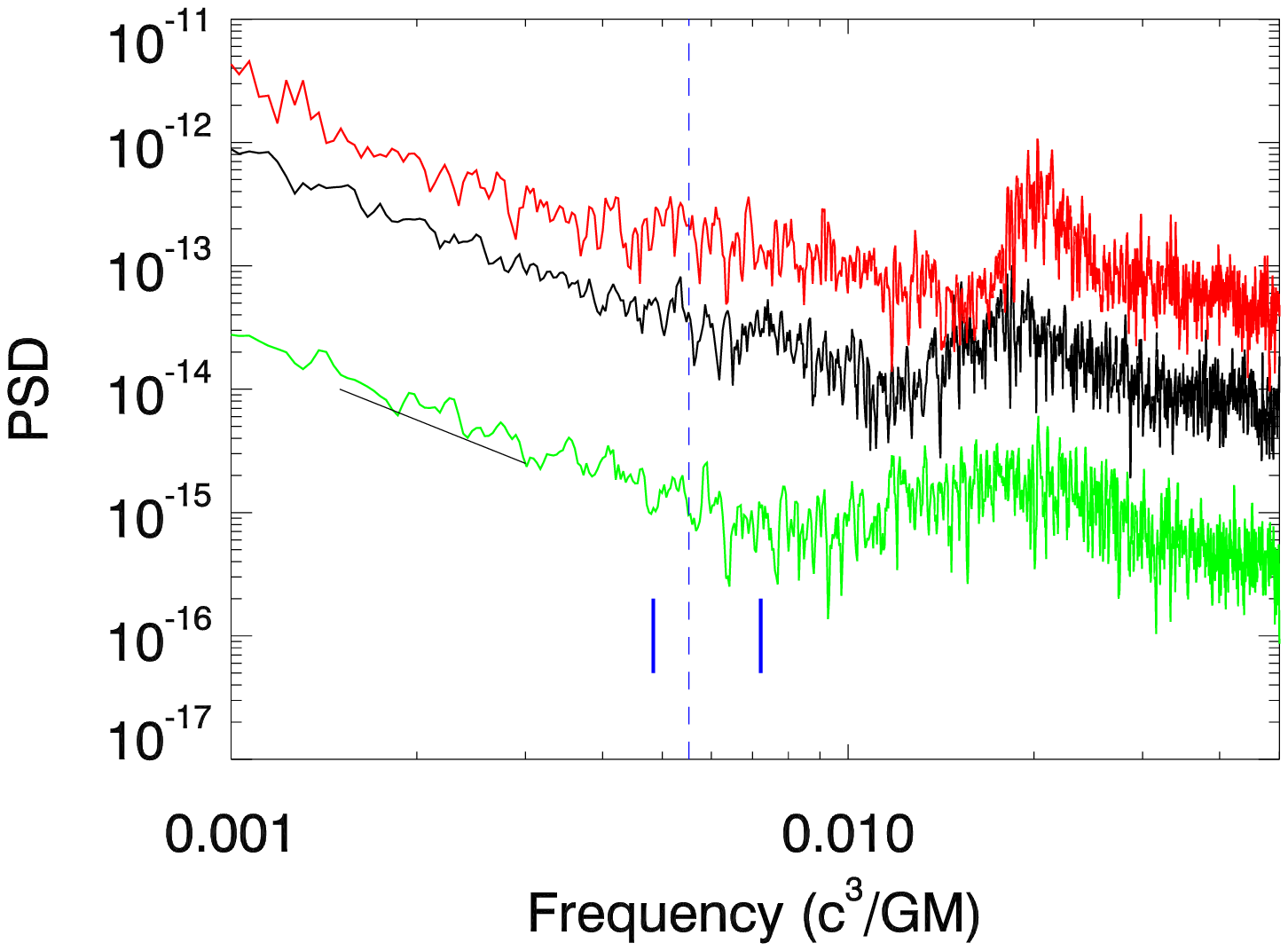,width=0.55\textwidth}
}
\caption{PSD of the midplane (decay-corrected) pressure for $\Delta
r=0.5r_g$ wide zones centered on $r=7r_g$ (top, red curve),
$r=8r_g$ (middle, black curve), and $r=9.2r_g$ (bottom, green
curve).  The thin vertical dashed line marks the maximum radial
epicyclic frequency.  A line-segment with a slope of $-2$ is shown for
reference.  Also shown (heavy vertical blue lines) are the radial and
vertical epicyclic frequencies at $r=9.2r_g$ where the simplest
form of the parametric instability model would predict resonances.}
\label{fig:mhd_1dpsd}
\vspace{0.5cm}
\end{figure}

In stark contrast to the hydrodynamic simulation, the
(decay-corrected) midplane pressure PSD does not show the
$\kappa$-bounded vertical ``ridge'' on the frequency-radius plane that
is characteristic of trapped g-modes.  The absence of an excited
trapped g-mode is also confirmed by examining the PSD at $r=8r_g$
(Fig.~\ref{fig:mhd_1dpsd}, normalized such that a direct comparison
can be made with the 3-d hydrodynamic results of
Fig.~\ref{fig:hd3d_psd}).  It is important to note, however, that a
trapped g-mode of the strength seen in our 3-d hydrodynamic simulation
would not stand out from the background turbulence.  Thus, while it is
apparent that the fundamental trapped g-mode is not strongly excited
by the turbulence (in the way that the local p-modes are, for
example), we cannot say whether the g-mode is actively damped by the
turbulence.

Finally, we note that there is no indication that the parametric
resonance instability of Abramowicz \& Kluzniak is at work, at least
in its simplest form.  As discussed in \S~2, this model predicts its
strongest resonance at the location where the radial and vertical
epicyclic frequencies are in a 2:3 ratio; this occurs at $r=9.2r_g$
in the PW potential.  As shown in Fig.~\ref{fig:mhd_1dpsd}, the PSD of
the pressure fluctuations at $r=9.2r_g$ shows no structure
associated with the local epicyclic frequencies.  We have verified
that a similar conclusion holds true for the PSD of the other
variables.  While it is possible that non-linear effects have shifted
the location of the resonance inwards from $r=9.2r_g$ (A2003), we
note that the PSDs of Fig.~\ref{fig:mhd_psd} show no obvious radius at
which the power at the radial epicyclic and the orbital frequencies
appears to be locally enhanced.

\subsection{Temporal properties of the instantaneous black hole accretion rate}
\label{sec:lightcurves}

The previous section addressed the temporal properties of the
fundamental fluid variables through the body of the disk.  Of course,
real observations of accretion disks measure the electromagnetic
radiation generated by the accretion flow.  While our simulations
miss all of the physics relevant for making a first-principles
prediction of the observed lightcurve, it is still instructive to
analyze a simple scalar quantity that can be generated from the
simulation and may be related to the observed hard X-ray radiation
(since it is the hard X-rays that carry the HFQPO signal).

\begin{figure}[b]
\centerline{
\psfig{figure=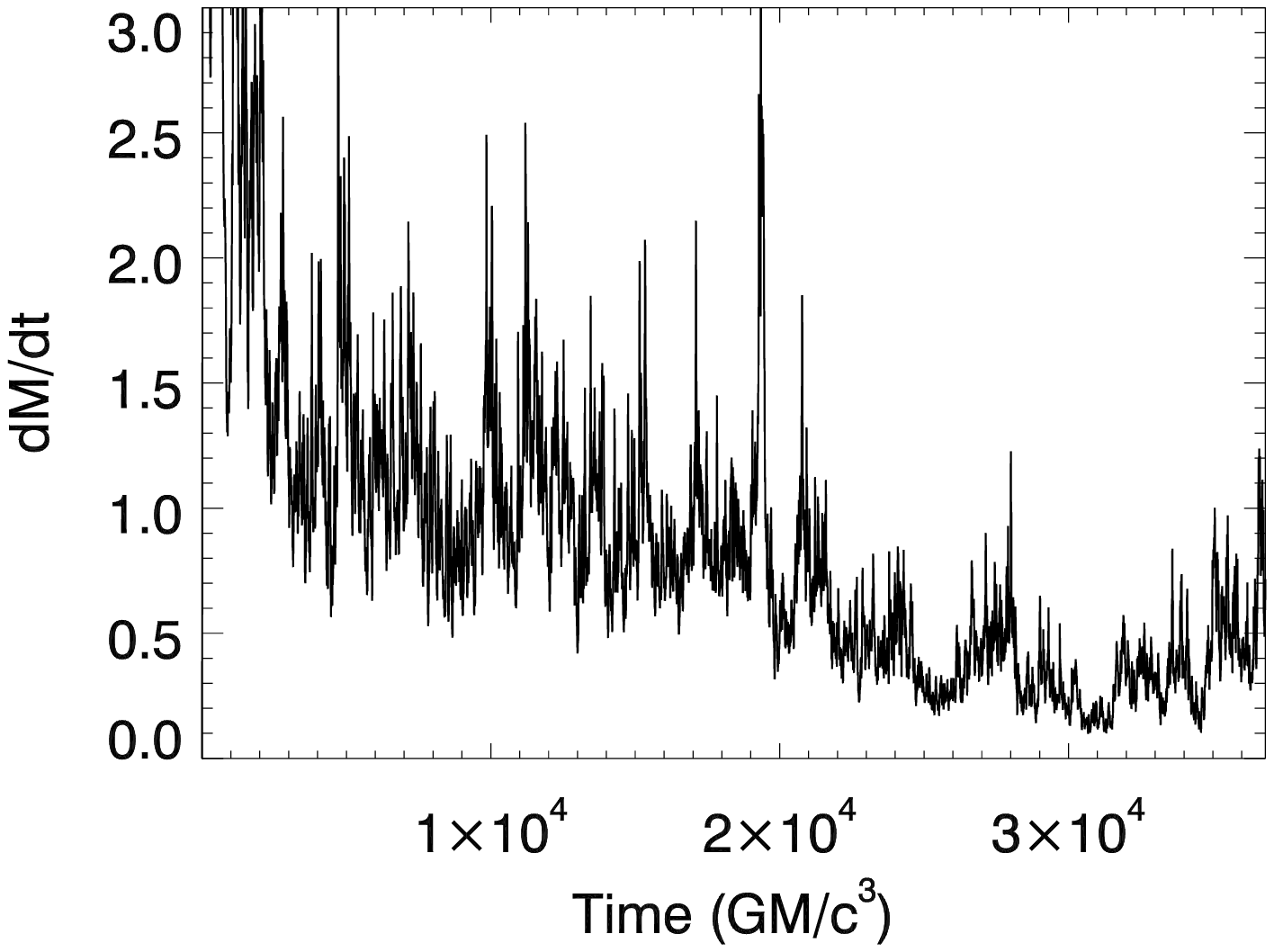,width=0.55\textwidth}
}
\caption{The instantaneous mass accretion rate onto the black hole
from our canonical MHD simulation (run MHD\_1).}
\label{fig:lightcurve}
\end{figure}

Here we consider one such proxy for the observed lightcurve, the
{\it instantaneous mass accretion rate} into the black hole,
\begin{equation}
\dot{M}=\int_{\partial {\cal R}\,_i}r(-v_r)\rho\,dS,
\end{equation}
where the integral is performed over the surface $\partial {\cal
R}\,_i$ defining the inner radial boundary of the computational
domain.  Figure~\ref{fig:lightcurve} shows the raw lightcurve (i.e.,
not corrected for the exponential decay of the disk).  In the
remainder of this section we address the temporal properties of this
lightcurve.  Of particular interest is the presence of breaks or QPOs
in the PSD of this light curve.  Hence, we need a quantitative
approach by which the significance of such features in the PSD can be
assessed.  We begin by discussing our general statistical approach,
which differs from the one advocated by SKH.

\subsubsection{Analysis method}

Our approach to PSD fitting is predicated on the assumption (also made
by SKH) that at a given frequency the power density has an exponential
probability distribution; if the mean is $p_0$, then the probability
of measuring a power between $p$ and $p+dp$ is
\begin{equation}
P(p)dp={1\over p_0}e^{-p/p_0}dp\; .
\end{equation}
Suppose that we have a model that predicts a power density
$p_{\rm mod,i}$ for frequency bin $i$, and that in our MHD
simulation we actually observe a power density $p_{\rm obs,i}$
in that bin.  The likelihood of the data given the model in
that bin is then
\begin{equation}
{\cal L}_i=(1/p_{\rm mod,i})\exp(-p_{\rm obs,i}/p_{\rm mod,i}).
\end{equation}
The likelihood of the whole power density spectrum given the
model is the product of the individual likelihoods, but the
log likelihood is typically more useful:
\begin{equation}
\ln{\cal L}=\sum_i \left[-\ln p_{\rm mod,i}-p_{\rm obs,i}/p_{\rm mod,i}
\right]\; .
\end{equation}
This is the figure of merit for a given model.  It can therefore be
used for standard tasks such as parameter estimation and model
comparison (e.g., determining if a QPO or break is required by the
data).  

Note that to maximize the information content, the bin size should be
the smallest possible, in this case the frequency resolution of the
raw PSD.  As a result, this method does not require rebinning to
coarser resolution.  Our approach therefore yields an accurate
evaluation of one or more precisely specified models, as opposed to
the broader but less sensitive method of trying to detect a signal in
a model-independent way.

We apply a Markov Chain Monte Carlo method to search for best fits and
establish confidence regions.  As discussed in
\S~\ref{sec:decay_corr}, we correct for secular changes.  Given the
large amplitude fluctuations in the $\dot{M}$ and $S_{\rm tot}$
curves, the end-point method discussed in \S~\ref{sec:decay_corr} is
not appropriate.  Hence, we employ the least-square method described
in \S4.3.1 in order to determine and then divide out the best-fitting
exponential decay.  

\subsubsection{Results}
\label{sec:mdot_psd}

\begin{figure*}
\hbox{
\psfig{file=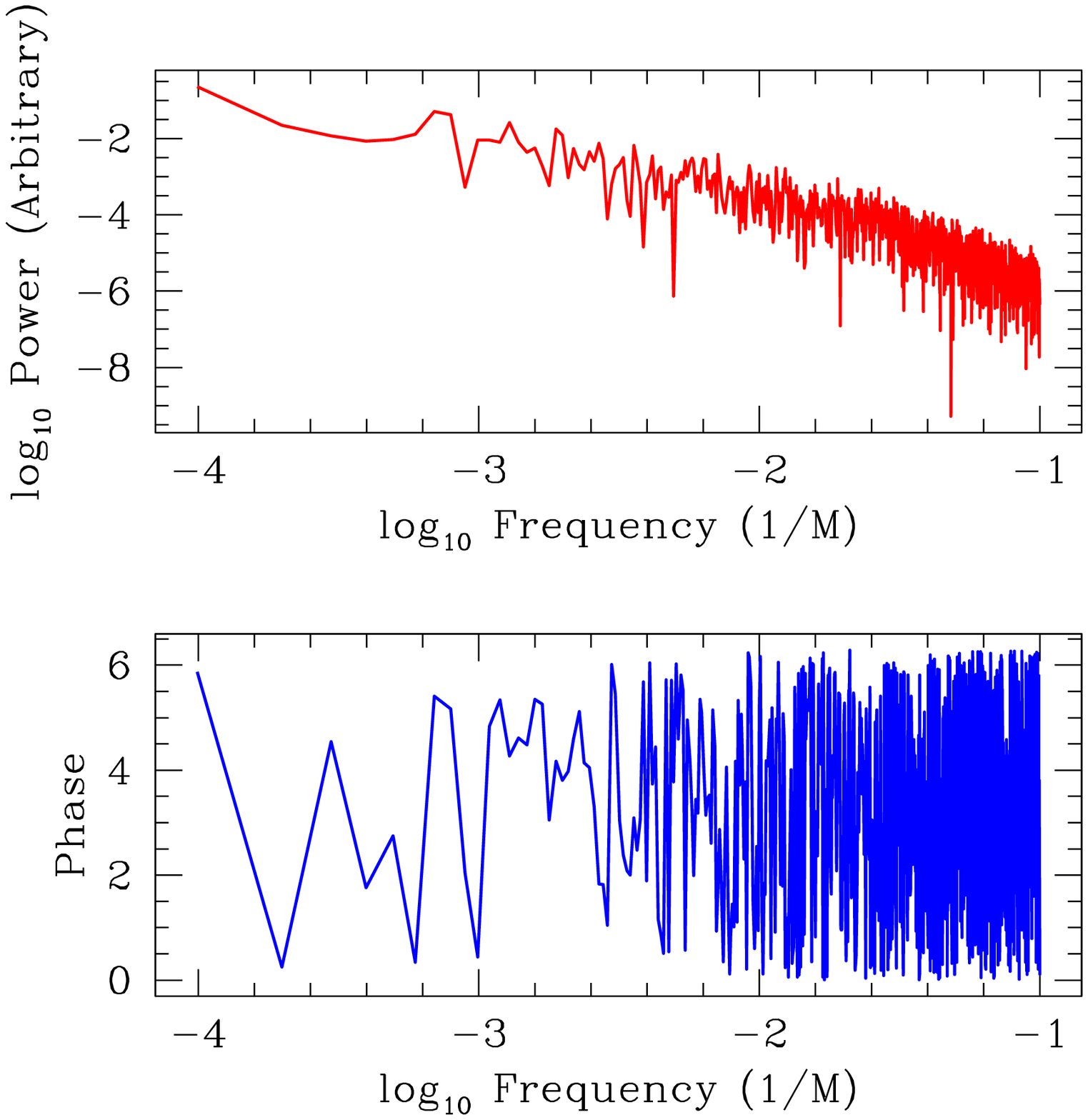,width=0.45\textwidth}
\psfig{file=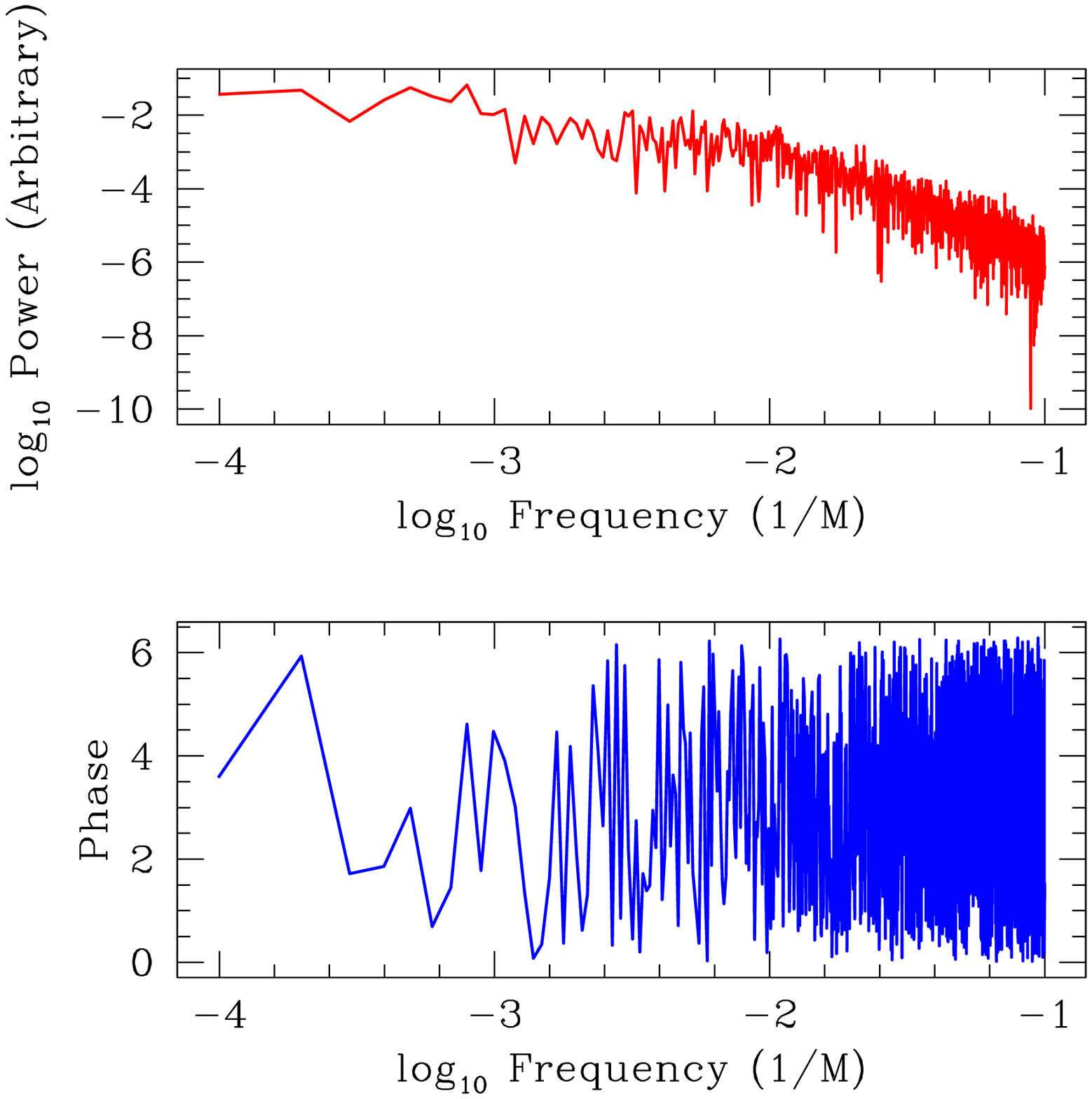,width=0.45\textwidth}
}
\caption{ Comparison of the PSD and Fourier phase of the mass
accretion rate without (left panel) and with (right panel)
multiplication of the time series by an exponential in time designed
to compensate for the slow loss of mass from the disk.  The similarity
of the two implies that at these frequencies the variability is
dominated by intrinsic fluctuations instead of by secular changes in
the disk.}
\label{fig:mdotcomp}
\end{figure*}

The mass accretion rate $\dot{M}$ has sufficient intrinsic variability
that secular changes in the disk properties have only minor effects on
the PSD.  This is evident from Fig.~\ref{fig:mdotcomp}, which compares
the PSD and Fourier phase of the raw $\dot{M}$ curve as a function of
frequency (left panel) with these after multiplication by the
exponential in time that minimizes the overall rms amplitude (right
panel).  The lack of significant differences suggests that inferences
drawn from the PSD are robust.  The comparison shown in
Fig.~\ref{fig:mdotcomp} is for the second quarter of run MHD\_1
(approximately $t=150T_{\rm isco}$ to $t=310T_{\rm isco}$); analyses
of the third and fourth quarters also reveal a lack of sensitivity to
the secular decay of the disk.

Having established that the secular decay of the disk is unimportant
for the PSD of ${\dot M}$, we will work directly with the raw time
series from run MHD\_1, without multiplication by an exponential
function.  We analyze separately the second, third, and fourth quarter
of the data (each encompassing a period of $\Delta t\approx 160T_{\rm
isco}$) to look for trends or stability in the PSD, while discarding
the first quarter as potentially biased by initial conditions.  The
results are summarized in Table~\ref{tab:tabmdot}, where we compare
single power law models for the PSD with models involving a broken
power law.  We use the method described in the previous section; note
that only differences in the log likelihood $\ln{\cal L}$ are
important rather than the absolute magnitude of $\ln{\cal L}$.

\begin{table*}
\centering
\begin{tabular}{lccr}\hline
Data segment&Model&Parameters and uncertainties&Maximum $\ln{\cal
L}$\\\hline
1&Single PL&$\Gamma=2.50\pm 0.05$&9131\\
&Broken PL&$\Gamma_1=0.82\pm 0.09$, $\Gamma_2=2.96\pm 0.07$,&9238\\
&&$\log_{10}(\nu_{\rm break})=-1.94\pm 0.04$\\\hline
2&Single PL&$\Gamma=2.70\pm 0.08$&6508\\
&Broken PL&$\Gamma_1=1.45\pm 0.11$, $\Gamma_2=2.91\pm 0.10$,&6609\\
&&$\log_{10}(\nu_{\rm break})=-1.65\pm 0.07$\\\hline
3&Single PL&$\Gamma=2.16\pm 0.04$&9561\\
&Broken PL&$\Gamma_1=1.47\pm 0.07$, $\Gamma_2=2.89\pm 0.09$,&9630\\
&&$\log_{10}(\nu_{\rm break})=-1.60\pm 0.04$\\\hline
\end{tabular}
\caption{Model comparisons for PSD of mass accretion rate $\dot{M}$.
$\Gamma$ is the power law index of a single power-law model for the
PSD; $P(\nu)\propto \nu^{-\Gamma}$.  $\Gamma_1$ and $\Gamma_2$ are
the
two indices obtained from a broken power law; $P\propto
\nu^{-\Gamma_1}$ for $\nu<\nu_{\rm break}$, $P\propto
\nu^{-\Gamma_2}$
for $\nu>\nu_{\rm break}$.  All error bars are one standard
deviation.}
\label{tab:tabmdot}
\end{table*}

From this analysis, we find compelling evidence for a break in the
power law characterizing the PSD of ${\dot M}$.  Compared to a single
power law, the broken power fit is better by $\Delta\ln{\cal
L}=76-90$, hence the maximum likelihood ratio is at least
$\exp(76)=10^{33}$ in all three data segments independently.  The
break frequency and power law slopes are consistent between the second
and third data segments, but these do not match the first data
segment.  The break frequencies are within a factor of two of, but not
consistent with, the orbital frequency at the ISCO, $\log_{10}(\nu_{\rm
ISCO})=-1.79$.  Therefore, although there is a clear steepening in the
power density spectrum, it is not possible at this point to assign a
specific physical meaning to the break.  We note that this general
form of the PSD, i.e., an approximate power-law with curvature or a
break at frequencies close to the ISCO orbital frequency, has been
previously seen in the mass accretion rate of global disk simulations
(Hawley \& Krolik 2001, 2002).

We see no indications of QPOs in the $\dot{M}$ PSD.  Quantitatively,
we add a Lorentzian QPO to the broken power-law PSD model in which the
QPO centroid frequency, full-width half-maximum, and amplitude are
allowed to be free parameters, with the one restriction that the
quality factor of the QPO must exceed 2.  We find that the peak power
of any QPO cannot exceed 2\% of the continuum power measured at the
centroid of the QPO at the 99\% confidence level.  

\section{Discussion}

\subsection{Comparison with previous numerical results}

In recent years, several groups have reported temporal analyses of MHD
disk simulations.  Here, we briefly compare our work with some of
these published results.

Probably the most relevant previous work is that of Arras, Blaes \&
Turner (2006; hereafter ABT).  These authors perform a local, shearing
box MHD simulation of a patch of an accretion disk; this provides a
controlled environment in which fluid modes can be characterized.  ABT
find that the MHD turbulence excites a spectrum of distinct acoustic
modes as well as radial epicyclic motions.  However, they note a lack
of distinct inertial modes (g-modes) and use this fact to argue
against the excitation of trapped g-modes in global accretion disks.
Our findings are completely in line with those of ABT, and represent
an extension of ABT's conclusions to global simulations of thin
accretion disks.

There have been QPOs reported from global simulations.  Kato (2004d)
performed a global MHD disk simulation in a PW potential and presented
an analysis of quantities derived from the mass accretion rate.
Through visual inspection of the resulting PSDs from four periods of
the (long) simulation, he reports a pair of transient QPOs and a pair of
QPOs that are labeled as persistent (although it is not clear that they
are present in the PSD of all data segments, and the statistical
significance of the features is unclear).  The QPOs are attributed
to resonances between the vertical and radial epicyclic frequencies,
and it is found that these QPO pairs have frequency ratios that are
{\it approximately} 3:2.  We find no evidence of these resonances in
our simulations.  In another interesting difference, an inspection of
the radially-resolved PSDs in Kato (2004d) reveal no signs of the
local p-modes that seem to feature so prominently in our PSDs.  While
the reason for these discrepancies is unclear, we do note that the
Kato (2004d) simulations have an order of magnitude less resolution in
both the azimuthal and (more importantly) the vertical direction,
although his simulations do have a significantly larger computational
domain.  It is possible that the Kato (2004d) simulations have failed
to adequately resolve the vertical dynamics of the thin disk.

Chan et al. (2006) have also performed global MHD disk simulations in
a PW potential using the pseudo-spectral algorithm of Chan, Psaltis,
\& \"Ozel (2005, 2006).  Their study included a post-processing of the
simulation to include detailed radiative transfer, and was explicitly
targeted at understanding the variability (including the large
amplitude flaring) of the hot accretion flow around the black hole at
the center of the Galaxy.  They found that the turbulence of the
quiescent flow could only produce a factor of two modulations in the
observed luminosity.  To model the large amplitude flares, they
introduce large density perturbations into the flow.  After being
perturbed, the disk displays a QPO with a frequency equal to the
orbital frequency at the magnetosonic point.  Given that the Chan et
al. study is exploring a rather different regime of accretion than our
present study (i.e., hot, thick accretion flows versus thin, cold
accretion flows), it is hard to make a direct comparison.

Finally, SKH have performed detailed analyses of General Relativistic
MHD (GRMHD) simulations of disks performed using the code of
De~Villiers \& Hawley (2003).  In particular, they have studied a long
($6000GM/c^3$) simulation of a disk around a Schwarzschild black hole,
focusing on the temporal behavior of proxy-lightcurves (rather than
the underlying fluid properties discussed in this paper).  It is
unclear from their analysis whether their simulated accretion disk has
excited local p-modes of the type that we find in this current work.
SKH do find, though, that a proxy-lightcurve based on radiative
transfer through the disk assuming black body emission and free-free
absorption displays QPOs with an approximate 3:2 ratio.  However,
these QPOs are transient, only appearing at certain times and certain
viewing inclinations.  

\subsection{Comments on 3:2 frequency ratios}

As described above, several authors have reported transient QPO pairs
from MHD simulations with frequency ratio 3:2.  It is tempting to
interpret these as resonance phenomena.  Here, we note that there are
several ways that approximate 3:2 ratios can be generated that do not
necessarily involve resonances and, hence, one should guard against
over-interpreting QPO pairs.  Indeed, we must remember that some
sources have QPO frequency ratios that are inconsistent with 3:2
(e.g., the 67~Hz and 41~Hz QPOs from GRS~1915+105; see Strohmayer
2001).

For example, consider the local vertical p-modes of the disk
(discussed in \S~\ref{sec:local_osc}).  At a given radius, the $n=1$
vertical pressure mode has a frequency of $\omega_{\rm
vert,1}=\sqrt{\gamma+1}\,\Omega$, where $\Omega$ is the orbital
frequency.  For a gas pressure dominated disk in which $\gamma=5/3$,
$\omega_{\rm vert,1}/\Omega=1.63$, and for a radiation pressure
dominated disk in which $\gamma=4/3$, $\omega_{\rm
vert,1}/\Omega=1.53$.  If some unspecified physical process enhances
emission from a given ring of the disk, one can imagine a situation
where QPOs are generated at the orbital frequency and the $n=1$
vertical pressure mode, thereby giving frequency ratios compatible
with the measured ratios in several sources.  Alternatively, the next
lowest vertical mode that has a vertical velocity node in the midplane
(and thus maximum variation of pressure and density there) has a
frequency of $\omega_{\rm vert,3}=\sqrt{3\gamma+1}\,\Omega$.  As a
result, when $\gamma=5/3$ we have $\omega_{\rm vert,3}/\omega_{\rm
vert,1}=1.5$, and when $\gamma=4/3$ the ratio is 1.46.  Once again, if
nature picks out a specific radius and the emission is modulated by
the $n=1$ and $n=3$ modes, we would see QPOs with frequency ratios
entirely consistent with the observations.

As another example, suppose that the disk emission is modulated at the
vertical and radial epicyclic frequencies, and that the emission is
distributed in radius according to a standard Page \& Thorne (1974)
disk.  The radial distribution of the emission then peaks close to the
radius where the radial epicyclic frequency is a maximum (and hence is
slowly changing with radius).  One might then expect to see a pair of
QPOs corresponding to these two epicyclic frequencies.  The frequency
ratio depends only weakly on the spin parameter, ranging from 1.46 at
$a/M=0$ to 1.7 at $a/M=0.9$.

\section{Conclusions}

The origin of HFQPOs remains elusive.  Our simulations of
geometrically-thin accretion disks have shown that MRI-driven MHD
turbulence does not excite the trapped g-modes of Nowak \& Wagoner
(1992), even when those modes definitively exist in the equivalent
hydrodynamic disk.  We have also shown that MHD turbulence does not
excite the parametric resonance instability of Abramowicz \&
Klu\'zniak (2001).  Instead, the only distinct modes found in our
simulated MHD disks are local vertical p-modes and radial epicyclic
oscillations.

Clearly, the failure of all simulations to date to produce stable QPO
pairs of the type seen in GBHBs suggests that either the QPOs are
too weak to be detected in the simulations or the models are missing
some important physical ingredients.  It is an open question whether
the global disk modes or the parametric instabilities discussed in
this paper can be excited once one includes the full effects of GR
close to rapidly spinning black holes and/or radiation physics.
Indeed, the fact that HFQPOs are only seen in rather special spectral
states (the soft intermediate state; e.g., see Belloni 2006) when the
accretion rate is thought to be comparable to the Eddington limit
suggests that radiation physics, in particular, may well be important
to the HFQPO mechanism.  It is also noteworthy that the transition
from the hard intermediate state into the soft intermediate state is
associated with powerful relativistic ejection events.  Thus, another
interesting possibility is that HFQPO production occurs in the black
hole magnetosphere (i.e., the base of the jet) and not the accretion
disk at all.

\section*{Acknowledgments}

We thank Eve Ostriker for insightful discussions as well as allowing
us the use of her MPI-parallelized version of ZEUS.  We are also
grateful to Andy Fabian, Gordon Ogilvie, Sean O'Neill, Jim Pringle,
Jon Miller, Phil Uttley, Simon Vaughan, and Bob Wagoner for comments
and discussion that significantly improved this paper.  All
simulations described in this paper were performed on the Beowulf
cluster (``The Borg'') supported by the Center for Theory and
Computation (CTC) in the Department of Astronomy at the University of
Maryland College Park.  CSR and MCM thank the National Science
Foundation for support under grant AST~06-07428.  CSR also gratefully
acknowledges the University of Maryland's Graduate Research Board
Semester Award Program which supported the early phases of this work.

\section*{References}

\noindent Abramowicz, M.~A., Almergren, G.~J.~E., Klu\'zniak, W.,
Thampan, A.~V., \& Wallinder, F. 2002, Class. Quant. Grav., 19, L57

\noindent Abramowicz, M.~A., Karas, V., Klu\'zniak, W., Lee, W.~H., \&
 Rebusco, P. 2003, PASJ, 55, 467 (A2003)

\noindent Abramowicz, M.~A., \& Klu\'zniak, W. 2001, A\&A, 374, L19

\noindent Abramowicz, M.~A., \& Klu\'zniak, W. 2003, Gen. Rel. Grav.,
35, 69

\noindent Arras P., Blaes O.~M., \& Turner N.~J. 2006, ApJ, 645, L65 (ABT)

\noindent Balbus, S.~A., \& Hawley, J.~F. 1991, ApJ, 376, 214

\noindent Balbus, S.~A., \& Hawley, J.~F. 1998, Rev. Mod. Phys., 70, 1

\noindent Belloni, T. 2006, Ad. Space Res., 38, 2801

\noindent Binney, J., \& Tremaine, S. 1987, Galactic Dynamics (Princeton:
Princeton Univ. Press), 359

\noindent Chan, C.~K., Psaltis, D., \& \"Ozel, F. 2005, ApJ, 628, 353

\noindent Chan, C.~K., Psaltis, D., \& \"Ozel, F. 2006, ApJ, 645, 506

\noindent Chan, C.~K., et al. 2006, ApJ, submitted (astro-ph/0611269)

\noindent Churazov, E., Gilfanov, M., \& Revnivtsev, M. 2001, MNRAS, 
321, 759

\noindent Cowling, T.~G. 1957, Quart. J. Mech. Appl. Math., 10, 129

\noindent De~Villiers, J.-P., \& Hawley J.~F. 2003, ApJ, 592, 1060

\noindent Friedman, J.~L., \& Schutz, B.~F. 1978, ApJ, 221, 937

\noindent Gammie, C.~F., McKinney, J.~C., \& T\'oth, G. 2003,
ApJ, 589, 444

\noindent Gierlinski M., Middleton M., Ward M., Done C., 2008, Nature,
in press

\noindent Gleissner T., Wilms J., Pottschmidt K., Uttley P., Nowak
M.A., Staubert R., 2004, A\&A, 391, 875.

\noindent Hawley J.F., Krolik J.H., 2001, ApJ, 548, 348

\noindent Hawley J.F., Krolik J.H., 2002, ApJ, 566, 164

\noindent Hawley, J.~F., Gammie, C.~F., \& Balbus, S.~A. 1996, ApJ, 464, 690

\noindent Hantao, J., Burin, M., Schartman, E., \& Goodman, J. 2006, Nature,
444, 343

\noindent Kato, S. 1990, PASJ, 42, 99

\noindent Kato, S. 1991, PASJ, 43, 557

\noindent Kato, S. 1993, PASJ, 45, 219

\noindent Kato, S. 2004a, PASJ, 56, 559

\noindent Kato, S. 2004b, PASJ, 56, 905

\noindent Kato, S. 2004c, PASJ, 56, L25

\noindent Kato, S., \& Fukue, J. 1980, PASJ, 322, 377

\noindent Kato, S., \& Honma, F. 1991, PASJ, 43, 95

\noindent Kato, Y. 2004d, PASJ, 56, 931

\noindent Krolik, J. 1999, ApJ, 515, L73

\noindent Lehr, D., Wagoner, R.~V., \& Wilms, J., 2000, astro-ph/0004211

\noindent Lubow, S.~H., \& Pringle, J.~E. 1993, ApJ, 409, 360 (LP93)

\noindent Lyubarskii, Yu.~E. 1997, MNRAS, 292, 679

\noindent Markoff, S., Nowak, M.~A., Corbel, S., Fender, R.~P., \& Falcke, H.
2003, New AR, 47, 491

\noindent Markovi\'c, D., \& Lamb, F.~K. 1998, ApJ, 507, 316

\noindent McClintock, J.~E., \& Remillard, R.~A. 2003, astro-ph/0306213

\noindent McHardy, I.~M., K\"ording, E., Knigge, C., Uttley, P., \& Fender,
R.~P. 2006, Nature, 444, 730

\noindent Miller K.A., Stone J.M., 2000, ApJ, 534, 398

\noindent Nowak, M.~A., \& Wagoner, R.~V. 1991, ApJ, 378, 656 (NW91)

\noindent Nowak, M.~A., \& Wagoner, R.~V. 1992, ApJ, 393, 697 (NW92)

\noindent Nowak, M.~A., \& Wagoner, R.~V. 1993, ApJ, 418, 187

\noindent Nowak, M.~A., Wagoner, R.~V., Begelman, M.~C., \& Lehr, D.~E.
1997, ApJ, 477, L91

\noindent Okazaki, A.~T., Kato, S., \& Fukue, J. 1987, PASJ, 39,
457

\noindent Ortega-Rodriguez, M., Silbergleit, A.~S., \& Wagoner, R.~V.
2001, ApJ, 567, 1043

\noindent Paczynski, B., \& Wiita, P.~J. 1980, A\&A, 88, 23 (PW)

\noindent Page, D.~N., \& Thorne, K.~S. 1974, ApJ, 191, 499

\noindent Perez, C.~A., Silbergleit, A.~S., Wagoner, R.~V., \&
Lehr, D.~E. 1997, ApJ, 476, 589

\noindent Reynolds, C.~S., \& Fabian, A.~C., 2008, ApJ, in press

\noindent Rezzolla, L., Yoshida, S.'i., Maccarone, T.~J., \& Zanotti, O.
2003a, MNRAS, 344, L37

\noindent Rezzolla, L., Yoshida, S.'i., \& Zanotti, O. 2003b, MNRAS, 344, 978

\noindent Schnittman, J.~D., Krolik, J.~H., \& Hawley, J.~F. 
2006, ApJ, 651, 1031 (SKH)

\noindent Shakura N.I., Sunyaev R.A., 1973, A\&A, 24, 337

\noindent Shafee R., McKinney J.C., Narayan R., Tchekhovkoy A., Gammie
C.F., McClintock J.E., 2008, ApJL, in press (arXive:0808.0860)

\noindent Silbergleit, A.~S., Wagoner, R.~V., \& Ortega-Rodriguez, M. 
2001, ApJ, 548, 335

\noindent Stone, J.~M., \& Norman, M.~L. 1992a, ApJS, 80, 753

\noindent Stone, J.~M., \& Norman, M.~L. 1992b, ApJS, 80, 791

\noindent Strohmayer, T.~E. 2001, ApJ, 554, L169

\noindent Uttley, P., \& McHardy, I.~M. 2001, MNRAS, 323, L26

\noindent Uttley, P., McHardy, I.~M., \& Vaughan, S. 2005, MNRAS, 359, 345

\noindent Vaughan, S., \& Uttley, P. 2005, MNRAS, 362, 235

\noindent Vernaleo, J.~C., \& Reynolds, C.~S. 2006, ApJ, 645, 83

\noindent Wagoner, R.~V., Silbergleit, A.~S., \& Ortega-Rodriguez, M.
2001, ApJ, 559, L25

\appendix

\section{Fundamental g-mode frequency for very small sound speeds}

For very small sound speeds, there are simplifications that allow the
frequency of the fundamental trapped g-mode in a hydrodynamic
accretion disk to be obtained analytically.  Here we base our
analysis on the equations and formalism of NW92.

NW92 examine the linearized equations describing the behavior of the
scalar potential
\begin{equation}
\delta u\equiv \delta P/\rho,
\end{equation}
where $\delta P$ is the Eulerian variation in the pressure.  The
radial equation for the perturbation is
\begin{equation}
\omega^2 c_s^2 {\partial^2\delta u\over{\partial r^2}}=
-\left(\omega^2-\gamma\Upsilon\Omega^2\right)
\left(\omega^2-\kappa^2\right)\delta u\; ,
\label{eq:diff_eqn}
\end{equation}
where $\omega$ is the mode angular frequency, $\Omega$ is the orbital
angular frequency, $\kappa$ is the radial epicyclic angular frequency,
and $\gamma$ is the usual adiabatic index $\gamma=5/3$.   $\Upsilon$ is 
determined by the quantization condition
\begin{equation}
{A+1/2\over{(1-4B)^{1/2}}}=j+1/2\; ,
\label{eq:quantization}
\end{equation}
where $A=\Upsilon-\zeta$ and
\begin{equation}
B={\zeta\over\gamma}{(\omega^2-
\gamma\Upsilon\Omega^2)\over{\omega^2}}
\end{equation}
with $\zeta=(\gamma-1)/\gamma= 2/5$.  For the fundamental mode, we
have $j=0$ and eq.~\ref{eq:quantization} can be solved to obtain
\begin{equation}
\Upsilon=0.4{\Omega^2\over \omega^2}-0.2.
\end{equation}

Now we concentrate on low sound speeds, $c_s\ll 1$ (within this
Appendix, speeds will be given in units of $c$; frequencies in units
of $c^3/(GM)$; lengths in units of $r_g$).  This implies that the
mode frequency $\omega\approx\kappa_{\rm max}$, where $\kappa_{\rm
max}$ is the maximum radial epicyclic frequency.  In this limit, we
note that the factor ($\omega^2-\gamma\Upsilon\Omega^2$) within
eqn.~\ref{eq:diff_eqn} is close to constant with radius.  Let us
define $D\equiv -(\omega^2-\gamma\Upsilon\Omega^2)>0$, where we
evaluate $D$ by assuming $\omega=\kappa_{\rm max}$.

We also note that near the maximum, the radial epicyclic
frequency has a parabolic form, $\kappa^2=\kappa_{\rm max}^2-
E(r-r_{\rm max})^2$, where $r_{\rm max}$ is the radius at which
$\kappa=\kappa_{\rm max}$.  The differential equation then becomes
\begin{equation}
{\partial^2\delta u\over{\partial r^2}}=\left[
-{D\over c_s^2}\left(\kappa_{\rm max}^2/\omega^2-1\right)
+{D\over c_s^2}{E\over\omega^2}\left(r-r_{\rm max}\right)^2\right]\delta u\; .
\end{equation}
This has the form of a harmonic oscillator, so we try a solution
of the type
\begin{equation}
\delta u\propto e^{-{1\over 2}C(r-r_{\rm max})^2}
\end{equation}
meaning that
\begin{equation}
{\partial^2\delta u\over{\partial r^2}}=
\left[C^2(r-r_{\rm max})^2-C\right]e^{-{1\over 2}C(r-r_{\rm max})^2}\; .
\end{equation}
This yields the conditions
\begin{eqnarray}
C&=&{D\over c_s^2}\left(\kappa^2_{\rm max}/\omega^2-1\right)\\
C^2&=&{D\over c_s^2}{E\over\omega^2}\; .\\
\end{eqnarray}
Defining $x\equiv 1/\omega^2$, these two conditions can be combined
(eliminating $C$) to yield
\begin{equation}
{D^2\over c_s^4}\left(\kappa_{\rm max}^4x^2-2\kappa_{\rm max}^2x+1
\right)={D\over c_s^2}Ex\; .
\end{equation}
Solving for $x$ gives
\begin{equation}
x={1\over{2\kappa_{\rm max}^4}}\left[2\kappa_{\rm max}^2+c_s^2E/D
\pm\sqrt{4\kappa_{\rm max}^2c_s^2E/D+c_s^4E^2/D^2}\right]\; .
\end{equation}
Since $\omega<\kappa_{\rm max}$, we choose the positive sign.
For small $c_s$, the first term in the square root dominates,
and the lowest order in $c_s$ gives
\begin{equation}
x\approx {1\over{\kappa_{\rm max}^2}}\left(1+{c_s\over{\kappa_{\rm max}}}
\sqrt{E\over D}\right)\; .
\end{equation}
This implies finally that
\begin{equation}
{\kappa_{\rm max}-\omega\over{\kappa_{\rm max}}}=
{c_s\over{2\kappa_{\rm max}}}\sqrt{E/D}\; .
\end{equation}
This clearly demonstrates that the fractional difference of the mode
frequency from the radial epicyclic maximum is linear in $c_s$.  

Evaluating this expression for the PW potential, we obtain
$\kappa^2_{\rm max}=1.202\times 10^{-3}$ 
(at $r=r_{\rm max}=4+2\sqrt{3}=7.464$), and 
\begin{equation}
{\kappa_{\rm max}-\omega\over{\kappa_{\rm max}}}=2.269c_s\; .
\end{equation}
In contrast, the Nowak \& Wagoner pseudo-Newtonian potential, 
\begin{equation}
\Phi_{\rm NW}=-{1\over r}+{3\over r^2}-{12\over r^3}\; .
\end{equation}
yields $\kappa^2_{\rm max}=8.607\times 10^{-4}$ (at $r_{\rm max}=7.746$) and 
\begin{equation}
{\kappa_{\rm max}-\omega\over{\kappa_{\rm max}}}=5.621c_s\; .
\end{equation}
For both potentials, the analytic frequencies agree extremely well
with direct numerical solutions to eqn.~\ref{eq:diff_eqn}.

\end{document}